\documentclass{aa}
\usepackage{graphicx}
\usepackage{txfonts}
\usepackage{xcolor}
\usepackage[colorlinks=true, citecolor=blue]{hyperref}
\usepackage{natbib}
\usepackage{pdflscape}
\usepackage{siunitx}
\usepackage{xspace}

\bibpunct{(}{)}{;}{a}{}{,}

\newcommand{\figcita}[1]{Fig. \ref{#1}}
\newcommand{\tabcita}[1]{Table \ref{#1}}
\newcommand{\refe}[1]{{\hypersetup{hidelinks}\hyperref[fig_point]{\color{blue}#1}}}

\newcommand{\ngc}[1]{\object{NGC\,#1}}

\newcommand{\hcnthree}[0]{HCN(3--2)\xspace}

\newcommand{\hcopthree}[0]{HCO$^+$(3--2)\xspace}

\newcommand{\hcnone}[0]{HCN(1--0)\xspace}
\newcommand{\hcopone}[0]{HCO$^+$(1--0)\xspace}
\newcommand{\cotwo}[0]{CO(2--1)\xspace}
\newcommand{\coone}[0]{CO(1--0)\xspace}

\newcommand{\tkin}[0]{T_\text{K}}

\newcommand{\vgrad}[0]{\nabla v}
\newcommand{\sstar}[0]{\Sigma_\textrm{star}}
\newcommand{\ssfr}[0]{\Sigma_\text{SFR}}
\newcommand{\meanco}[0]{\langle I_\text{CO}\rangle}

\begin{document} 

\title{
Sub-kiloparsec empirical relations and excitation conditions of HCN and HCO$^+$ $J$=3-2 in nearby star-forming galaxies}

\author{
    A. Garc\'ia-Rodr\'iguez \inst{\ref{affil_oan}}
    \fnmsep
    \thanks{\email{axel.garcia@oan.es}}
\and
    A. Usero
    \inst{\ref{affil_oan}}
\and
    A. K. Leroy
        \inst{\ref{affil_osu}} 
\and  
        F. Bigiel
        \inst{\ref{affil_ubonn}}
\and
        M. J. Jim\'enez-Donaire
        \inst{\ref{affil_oan}}
\and
        D. Liu
        \inst{\ref{affil_mpe}}
\and
        M. Querejeta
        \inst{\ref{affil_oan}}
\and
        T. Saito
        \inst{\ref{affil_naoj}}
\and
        E. Schinnerer
        \inst{\ref{affil_mpia-heidelberg}}
\and
        A. Barnes
        \inst{\ref{affil_ubonn}}
\and
        F. Belfiore
        \inst{\ref{affil_inaf}}
\and
        I. Be\v{s}li\'c
        \inst{\ref{affil_ubonn}}
\and
        Y. Cao
        \inst{\ref{affil_umars}}
\and
        M. Chevance
        \inst{\ref{affil_ita},\ref{affil_cool}}
\and
        D. A. Dale
        \inst{\ref{affil_uwyoming}}
\and
        J. S. den Brok
        \inst{\ref{affil_ubonn}}
\and
        C. Eibensteiner
        \inst{\ref{affil_ubonn}}
\and
        S. Garc\'ia-Burillo
        \inst{\ref{affil_oan}}
\and
        S.~C.~O. Glover
        \inst{\ref{affil_ita}}
\and
        R. S. Klessen 
        \inst{\ref{affil_ita},\ref{affil_izfwr}}
\and
    J. Pety
    \inst{\ref{affil_iram},
    \ref{affil_usorbonne}}
\and
        J. Puschnig
        \inst{\ref{affil_ubonn}}
\and
        E. Rosolowsky
        \inst{\ref{affil_ualberta}}
\and
        K. Sandstrom
        \inst{\ref{affil_ucsd}}
\and
        M. C. Sormani
        \inst{\ref{affil_ita}}
\and
        Y.-H. Teng
        \inst{\ref{affil_ucsd}} 
\and
        T. G. Williams
        \inst{\ref{affil_mpia-heidelberg}}
}
          
\institute{
    Observatorio Astron\'omico Nacional (IGN), C/ Alfonso XII, 3, Madrid E-28014, Spain 
    \label{affil_oan} 
\and
    Department of Astronomy, The Ohio State University, 140 West 18th Ave, Columbus, OH 43210, USA
    \label{affil_osu}
\and
    Argelander-Institut f\"{u}r Astronomie, Universit\"{a}t Bonn, Auf dem H\"{u}gel 71, 53121 Bonn, Germany
    \label{affil_ubonn}
\and
    Max-Planck-Institut f\"{u}r extraterrestrische Physik, Giessenbachstra{\ss}e 1, D-85748 Garching, Germany
    \label{affil_mpe}
\and
	National Astronomical Observatory of Japan, 2-21-1 Osawa, Mitaka, Tokyo, 181-8588, Japan
	\label{affil_naoj}
\and
    Max Planck Institute for Astronomy, K\"{o}nigstuhl 17, 69117 Heidelberg, Germany
    \label{affil_mpia-heidelberg}
\and
	INAF – Osservatorio Astrofisico di Arcetri, Largo E. Fermi 5, I-50157 Firenze, Italy
	\label{affil_inaf}
\and
 	Aix Marseille Université, CNRS, CNES, LAM (Laboratoire d’Astrophysique de Marseille), F-13388 Marseille, France
 	\label{affil_umars}
\and
	Institut für Theoretische Astrophysik, Zentrum für Astronomie, Universität Heidelberg, Albert-Ueberle-Straße 2, D-69120 Heidelberg, Germany
	\label{affil_ita}
\and
    Cosmic Origins Of Life (COOL) Research DAO, coolresearch.io
    \label{affil_cool}
\and
	Department of Physics and Astronomy, University of Wyoming, Laramie, WY 82071, USA
	\label{affil_uwyoming}
\and
	Interdisziplinäres Zentrum für Wissenschaftliches Rechnen, Universität Heidelberg, Im Neuenheimer Feld 205, D-69120 Heidelberg, Germany
	\label{affil_izfwr}
\and
    IRAM, 300 rue de la Piscine, F-38406 Saint Martin d'H\`eres, France
    \label{affil_iram}
\and
    Sorbonne Universit\'e Observatoire de Paris, Universit\'e PSL, \'Ecole normale sup\'erieure, CNRS, LERMA, F-75005, Paris, France
    \label{affil_usorbonne}
\and
	4-183 CCIS, University of Alberta, Edmonton, AB T6G 2E1, Canada
	\label{affil_ualberta}
\and
	Center for Astrophysics and Space Sciences, Department of Physics, University of California San Diego, 9500 Gilman Drive, La Jolla, CA 92093, USA
	\label{affil_ucsd}
}

\date{Received --- --, 2020; accepted --- --, ----}
 
\abstract{
We present new HCN and HCO$^+$ ($J$=3--2) images of the nearby star-forming galaxies (SFGs) \ngc{3351}, \ngc{3627}, and \ngc{4321}. The observations, obtained with the Morita ALMA Compact Array, have a spatial resolution of $\sim$290--440 pc and resolve the inner $R_\textrm{gal} \lesssim$ 0.6-1~kpc of the targets, as well as the southern bar end of \ngc{3627}. We complement this data set with publicly available images of lower excitation lines of HCN, HCO$^+$, and CO and analyse the behaviour of a representative set of line ratios: \hcnthree/\hcnone, \hcnthree/\hcopthree, \hcnone/\cotwo, and \hcnthree/\cotwo. Most of these ratios peak at the galaxy centres and decrease outwards. We compare the HCN and HCO$^+$ observations with a grid of one-phase, non-local thermodynamic equilibrium (non-LTE) radiative transfer models and find them compatible with models that predict subthermally excited and optically thick lines. We study the systematic variations of the line ratios across the targets as a function of the stellar surface density ($\sstar$), the intensity-weighted \cotwo ($\meanco$), and the star formation rate surface density ($\ssfr$). We find no apparent correlation with $\ssfr$, but positive correlations with the other two parameters, which are stronger in the case of $\meanco$. The HCN/CO-$\meanco$ relations show $\lesssim$0.3~dex galaxy-to-galaxy offsets, with \hcnthree/\cotwo-$\meanco$ being $\sim$2 times steeper than \hcnone/\cotwo. In contrast, the \hcnthree/\hcnone-$\meanco$ relation exhibits a tighter alignment between galaxies. We conclude that the overall behaviour of the line ratios cannot be ascribed to variations in a single excitation parameter (e.g. density or temperature). 
}


\titlerunning{Sub-kpc excitation conditions in nearby star-forming galaxies}
\authorrunning{Garc\'ia-Rodr\'iguez et al.}

\maketitle

\section{Introduction}
\label{sec_intro}

Gas volume density ($n$; hereafter, density) is a key parameter of star formation (SF) theories since it regulates the onset and timescale of the gravitational collapse of molecular clouds \citep{mckee2007,klessen2016}. Unfortunately, density is also an elusive quantity from an observational point of view. In the last decades, the $J$=1--0 lines of HCN and HCO$^+$ have been widely adopted as proxies for the dense molecular phase in external galaxies \citep{hel97,gao04sfr,gra08}. Compared with the low-$J$ lines of CO (the default bulk gas tracer), HCN and HCO$^+$ lines have $\sim$30-40 times stronger dipoles \citep{sch05} and lower opacities, thus requiring $\sim$1-2 dex higher densities to be effectively excited. This would favour the HCN and HCO$^+$ emission being dominated by dense gas close the phase that is actually forming stars. Based on this assumption, several surveys have derived constraints on SF theories by studying how the CO and HCN (or HCO$^+$) emission correlates with the star formation rate (SFR) within galaxies 
\citep{use15,gal18,jim19} or among different galaxy populations \citep{gao04sfr,gra08,bur12}. 

The assumption that HCN and HCO$^+$ lines are tracers of dense gas mass has been questioned by some studies of Galactic clouds (\citealt{kau17,pet17,bar20,taf21}). In agreement with recent simulations \citep{jon21}, those studies have found that the total \hcnone and \hcopone luminosities of some clouds have significant contributions from gas at moderate or even low volume densities (${\sim} 10^{2-3}$ cm$^{-3}$). This can be explained by a decrease in the effective density for excitation due to radiative trapping \citep{sco74}, and by the fact that density PDFs of molecular clouds are bottom-heavy.  

It is thus mandatory to revise the excitation of the \hcnone and \hcopone lines in an extragalactic context. The HCN or HCO$^+$ $J$=3--2 lines are ideal tools for this purpose since their fiducial critical densities for excitation are $\gtrsim$1~dex higher \citep{shi15}, thus reducing the contribution from low-density gas. Due to their faintness, observations of the $J$=3--2 lines in external galaxies have been relatively scarce and mostly limited to bright sources \citep[e.g.][]{gra08,sai18,tan18}. Very few works have observed them in normal, star-forming galaxies (SFGs; L$_\textrm{TIR} < 10^{11} \, \textrm{L}_\odot$), let alone resolving their spatial distributions. Dedicated $J$=3--2 observations can help to understand the general behaviour of the $J$=1--0 lines, which, regardless of the ambiguities pointed out by Galactic studies, remain more accessible in nearby galaxies.

In this series of papers, we report on 290--440~pc resolution images of the \hcnthree and \hcopthree emission in three SFGs obtained with the Morita Atacama Compact Array (ACA). This first paper describes the new data, which we compare with previous interferometer images of lower-excitation lines (\hcnone, \hcopone, \coone, and \cotwo). We use this data set to resolve and discuss the excitation of HCN and HCO$^+$ across the discs of SFGs for the first time. In a subsequent paper (hereafter Paper~II), we study the distribution of the molecular interstellar medium (ISM) properties across the three galaxies by means of a detailed multi-zone radiative transfer modelling.

\section{Sample selection}\label{sec_sample}

\begin{table*}
\centering
\caption{Main properties of our targets.}
\begin{tabular}{llccccccccc}
\hline
\hline
NGC & Morph. & RA$_\text{J2000}$ & DEC$_\text{J2000}$ & $D$ & $M_\star$ & SFR & $i$ & $PA$ & $\varv_\mathrm{sys, LSR}$ & Scale at $6''$ \\
           & Type & (hh:mm:ss.s) & (\si{\degree}:\si{\arcmin}:\si{\arcsec}) & (Mpc) & ($10^{10}~M_\odot$) & ($M_\odot$~yr$^{-1}$) & (\si{\degree}) & (\si{\degree}) & (km~s$^{-1}$) & (kpc) \\ 
\hline
\noalign{\smallskip}
{3351} & SB(r)b     & 10:43:57.8 & 11:42:13.2 & 10.0 & 2.3 & 1.3 & 45 & 193 & \phantom{1}775 & 0.29 \\
{3627} & SAB(s)b   & 11:20:15.0 & 12:59:29.4 & 11.3 & 6.8 & 3.8 & 57 & 173 & \phantom{1}715 & 0.33 \\ 
{4321} & SAB(s)bc & 12:22:54.9 & 15:49:20.3 & 15.2 & 5.6 & 3.6 & 38 & 156 & 1572 & 0.44 \\
\hline
\end{tabular}
\tablefoot{Column description: Source name; morphological type \citep{vau91}; centre coordinates from \citet{lan20} (they coincide with the phase centre of the ACA observations, except for \ngc{3627}, whose phase centre was at [11:20:15.7, 12:59:09.9]); distance from \citet{jac09} (\ngc{3351} and \ngc{3627}) and \citet{fre01} (\ngc{4321}) \citep[for details, see][]{ana21}; total stellar mass and SFR \citep{ler21b}; inclination, position angle, and systemic velocity \citep{lan20}; and physical scale at our working resolution of 6$''$.}\label{tab_sample}
\end{table*}

The motivation for our \mbox{\hcnthree} and \hcopthree observations was to resolve the excitation of these species in SFGs, which has been poorly studied so far (Sect.~\ref{sec_intro}). To select the targets for the ACA, we searched the ALMA archive for nearby SFGs with available high-quality maps of lower-excitation lines (i.e. \hcnone, \hcopone, and low-$J$ CO).
We selected \ngc{3351}, \ngc{3627}, and \ngc{4321} (\tabcita{tab_sample}), which are the brightest objects in a small \hcnone and \hcopone survey by \cite{gal18}, belong to the PHANGS-ALMA sample \citep{ler21b}, and have a rich ancillary data set.

The targets are barred spiral galaxies with Milky Way-like stellar masses and SFRs, hosting negligible (\ngc{3351}) to weak (\ngc{3627}: LINER; \ngc{4321}: H{\sc ii}/LINER) nuclear activity \citep{ho97, gad19}. However, they show different gas morphologies over the inner disc regions studied here (Fig.~\ref{fig_mom0-HCN32}, bottom). \textbf{\ngc{3351}} hosts a prominent starburst ring \citep[SFR$\sim$0.6 M$_\odot$;][]{gad19,lin20} at $\sim$0.4 kpc radius, where dynamical resonances have piled up the gas at the two contact points with the bar \citep{lea19, wil21}.  \textbf{\ngc{3627}} hosts interaction-triggered gas inflows through the spiral arms and bar \citep{ile22} that enhance the star formation rate in the bar ends through gas compression \citep{mur15,wat19}. \textbf{\ngc{4321}'s} powerful bar is flanked by two spiral-like dust lanes
down to the nuclear region, where they join a $\sim$0.6~kpc-long secondary bar \citep{gar98,erw04} that is parallel to the main one. At our $\sim$290--440~pc resolution, some of these morphological features are partly blurred (e.g. compare colours and contours in Fig.~\ref{fig_mom0-HCN32}-bottom), but still recognisable in some line maps.

\section{Data and physical parameters}\label{sec_observations}

\subsection{Molecular line data}

\subsubsection{New ACA band-6 observations}
\label{sec_newACA}

\begin{table*}
\centering
\caption{Summary of the new ACA observations.}
\begin{tabular}{lcccccccc}
\hline
\hline
NGC & Obs. Time & Mosaic & No. EB & $(u,v)$ Lengths & MRS & Native Beam & RMS noise$^{\dag,\star}$ ($\sigma_\textrm{channel}$) & Jy/K$^\dag$ \\ 
            & (h) &  & & (m) & ($''$) & ($''\times''$) & (mK)  \\ 
\hline
\noalign{\smallskip}
{3351} & 5.0 & 1 pointing & 4 & 7.0--44.2 & 33 & $5.75 \times 4.12$& 2.0 & 2.07\\ 
{3627} & 8.6 & 3-point linear & 6 & 7.0--44.7 & 29 & $5.78 \times 4.23$ & 1.8 & 2.07\\
{4321} & 4.3 & 1 pointing & 3 & 7.0--44.5 & 34 & $5.88 \times 4.27$ & 2.1 & 2.06\\ 
\hline
\end{tabular}
\tablefoot{Column description: Galaxy name; Total observing time; Number of pointings; Number of execution blocks (EB); Total range of $(u,v)$ baseline lengths; maximum recoverable scale (MRS); native elliptical resolution of the synthesised beam; RMS noise of the cubes at the phase centre; Jansky-to-Kelvin conversion factor. Frequency-dependent properties correspond to the \hcnthree rest frequency.\\
$\dag$: measured at the final 6$''\times$10~km~s$^{-1}$ resolution. \\$\star$: measured at the phase centre.}\label{tab_obs}
\end{table*}

We obtained new band-6 (211--275~GHz) observations with the ACA in ALMA Cycle 6 under project 2018.1.01530 (PI: A. Usero). We simultaneously observed \hcnthree and \hcopthree ($\sim$265-267~GHz sky frequency) in the three target galaxies. Table~\ref{tab_obs} presents a summary of the observations. The data were acquired during April and May 2019, with 11 antennas available in all but a few cases. We observed a single pointing towards the inner discs of \ngc{3351} and \ngc{4321} -- the full width at half maximum (FWHM) of the primary beam at the observing frequency is $39''$ -- and a  three-pointing linear mosaic in \ngc{3627}. The footprints are shown in Fig.~\ref{fig_mom0-HCN32} (bottom). The spectral setup included two 1.875~GHz spectral windows centred on the redshifted \hcnthree and \hcopthree lines with a spectral resolution of 1.938~MHz (${\sim}2.2$~km~s$^{-1}$).

The raw ACA data were calibrated with the ALMA calibration pipeline implemented in the CASA software \citep[version 5.4.0-70;][]{mcmul07}. We only added a few minor flags when one of the antennas showed an unexpected amplitude or phase behaviour in the point-source calibrators. Next, we processed the calibrated measurement set (MS) files with the PHANGS-ALMA pipeline \citep[version 2.0; for details, see][]{ler21a}. The pipeline generated CLEANed, continuum-subtracted line cubes corrected for primary beam attenuation, expressed in Kelvin units, with a pixel size of $0.84''$, and a typical beam of $\sim 5.8 \si{\arcsec}\times4.2 \si{\arcsec}$. We confirmed that the output cubes had good quality and did not show apparent emission in the residuals. We also verified that the statistics of negative values is similar in channels with and without line emission, suggesting that the CLEANing -- performed on $\sim$8.8~km~s$^{-1}$ channels as a tradeoff between resolution and the signal-to-noise ratio (S/N) -- was as deep as possible given our sensitivity. As a final step, we brought all the data cubes to a common angular and spectral resolution using the GILDAS software package\footnote{\url{https://www.iram.fr/IRAMFR/GILDAS/}}. The elliptical Gaussian beam was rounded to 6\si{\arcsec} ($\approx$290--440~pc) by convolution with a Gaussian kernel. Velocity channels were bilinearly interpolated onto a 10 km~s$^{-1}$ grid. We truncated the images where the primary beam correction is $>$4. For each line cube, we generated a noise map by measuring the rms over channels outside the velocity range of the galaxy (Table~\ref{tab_obs}). 

We checked if the observations filter out a significant fraction of extended emission above their maximum recoverable scale (MRS; $\sim$30$\si{\arcsec}$ $\approx$1.6--2.4~kpc). We compared the fluxes of the PHANGS-ALMA \cotwo maps of our targets derived from ACA observations with and without short-spacing corrections at matching resolution. Since \cotwo, \hcnthree, and \hcopthree have similar rest frequencies, the beam sizes and MRS of the 7m-array observations are similar. We found that, on a pixel-by-pixel basis within our S/N-clipped regions (see Sect.~\ref{sec_generationMoments}), the 7m array alone missed $\sim$5\% (\ngc{3351}), $\sim$10\% (\ngc{3627}), and $\sim$18\% (\ngc{4321}) of the \cotwo flux recovered by the 7m+Total Power observations. These percentages only show small spatial variations. These are upper bounds for the percentage of flux filtered out by our 7m-only observations of \hcnthree and \hcopthree, since the distribution of the HCN and HCO$^+$ emission in galaxy discs is radially more compact than that of CO \citep[e.g.][]{gal18,jim19}. A related issue is that CLEAN might not recover faint emission that remains below the noise level in individual channels \citep{ler21a}. This could explain why some integrated intensity maps (Fig.~\ref{fig_mom0-HCN32}) show signatures of negative sidelobes (amplified by the primary beam correction), whereas the line cubes do not. The statistics of integrated intensity values points to a mean deficit at $\sim$1$\sigma$-level in \ngc{4321} and less significant in the other two galaxies. The impact of this on the spatially integrated flux could be important, but, on a pixel-by-pixel basis, it would remain $\lesssim$20\% for the S/N>5 regions that we select for analysis. Finally, we emphasise that, since the spatial distributions and S/N of \hcnthree and \hcopthree are similar, any flux loss would affect them in a very similar way. 

\subsubsection{Archival data}
\label{sec_archivalLineData}

\begin{table*}
\centering
\caption{Molecular lines studied in this paper.}
\begin{tabular}{lrcccccc}
\hline
\hline
\noalign{\smallskip}
\multicolumn{3}{c}{Lines} && \multicolumn{4}{c}{Observations}\\
\noalign{\smallskip}
\cline{1-3}\cline{5-8}
\noalign{\smallskip}
name & $\nu_{\rm rest}$ & $n_\text{eff, 20K}^\text{S15}$ &&
 $\theta_{B}$ & $\sigma_\text{channel}$ & telescope & ref. \\
\noalign{\smallskip}
  & (GHz) & (cm$^{-3}$) && (\si{\arcsec}) & (mK) \\
  \noalign{\smallskip}
\cline{1-3}\cline{5-8}
\noalign{\smallskip}
\hcnthree     & 265.89 & $7.3\,10^4$ && 6 & 1.8 -- 2.2 & ALMA (7m) &  1 \\
\hcopthree   & 267.56 & $6.8\,10^3$ && 6 & 1.8 -- 2.0 & ALMA (7m)  &  1 \\
\noalign{\smallskip}
\hcnone        &    88.63 &  $4.5\,10^3$                                     && 5 & 2.4 -- 5.0 & ALMA (12m) + IRAM30m & 2 \\
\hcopone      &    89.19 &  $5.3\,10^2$                                     && 5 & 4.2 -- 6.6 & ALMA (12m) + IRAM30m & 2 \\
\coone           & 115.27 &   $\hspace{3mm}\sim10^{2\dag}$    && 8 & --- & BIMA+NRAO12m or ALMA (12m+7m+TP) & 3 \\
\cotwo           & 230.54 &   $\hspace{3mm}\sim10^{2\dag}$    && 1.5-1.7 & 5.6 -- 16.7 & ALMA (12m+7m+TP) & 4 \\
\noalign{\smallskip}
\hline
\end{tabular}
\tablefoot{Column description: Line name; rest frequency; effective excitation density from \citet{shi15} at $\tkin=20$~K ($\dag$: CO lines are not included, so we quote a reasonable guess); native angular resolution of the data products; RMS noise per channel at the phase centre of each convolved and reprojected cubes ($6\si{\arcsec} \times 10$ km s$^{-1}$); telescope; and reference (see Sect.~\ref{sec_archivalLineData} for details). The latter are:
(1) this paper; 
(2) \citet{gal18} and \citet{jim19}; 
(3) \citet{hel03} for NGC~3351 and NGC~3627 and ALMA Verification Program for NGC~4321;
(4) \citet{ler21b}. }\label{tab_lineCatalogue}
\end{table*}

We complement our new ACA data with previous line observations of our three targets (\tabcita{tab_lineCatalogue}).  \citet{gal18} obtained $\sim$5$''$ resolution \textbf{\hcnone and \hcopone} images with the ALMA 12m array. They applied short-spacing corrections using IRAM 30m images from the EMPIRE survey \citep{jim19} and project 058-16 (PI: A.~Usero). \citet{gal18} also collected \textbf{\coone} images and homogenised them to a common angular resolution of $8''$. Those for \ngc{3351} and \ngc{3627} were fetched from the BIMA SONG survey \citep{hel03}. They combined BIMA interferometer observations with short-spacing information from the NRAO 12-m antenna at Kitt Peak. The \ngc{4321} image was part of the ALMA Science Verification program\footnote{\url{https://almascience.eso.org/alma-data/science-verification}} and combined observations with the 12m, 7m, and Total Power arrays. We obtained \textbf{\cotwo}  line cubes from the PHANGS-ALMA survey \citep[public data release -- version 4.0,][]{ler21b}. These data have high spatial resolution (1.5$''$--1.7$''\approx$71--121~pc) comparable to typical giant molecular cloud (GMC) scales, high sensitivity, and broad $(u,v)$ coverage from the combination of the 12m, 7m and Total Power arrays. Throughout this paper, we derive bulk molecular gas properties from these high-quality \cotwo data, rather than from the \mbox{\coone} images. 

All of these line cubes have a wider Field of View (FoV) than our ACA images. We convolved  them to our working resolution of $6''\times10$~km~s$^{-1}$, resampled them on the voxel grids of the ACA data, and scaled them to Kelvin units when necessary. In the case of the $8''$-resolution \coone data, we  scaled the \cotwo maps at 6$''$ by the \mbox{\coone/\cotwo} ratio maps at 8$''$ to generate artificial 6$''$-resolution \coone images. From a study of nearby galaxies at similar resolution (Saito et al., in prep.), we expect this approximation to be accurate within $\lesssim$5\%. Noise maps were generated as in Sect.~\ref{sec_newACA}.

As a reference, we quote in \tabcita{tab_lineCatalogue} the effective excitation densities defined by \citet{shi15} in terms of detectability (density needed to generate a 1~K~km~s$^{-1}$ line). Other definitions exist  \citep[e.g.][]{ler17} but these values already illustrate that the relative strength of the lines could vary across $\lesssim$3~dex in density. 

\subsubsection{Generation of moment maps}
\label{sec_generationMoments}

We generated moment maps by integrating the line cubes over a certain velocity window. Rather than fixing the same broad window across an entire galaxy, we adapted the window edges to the line emission within each pixel. This allowed us to recover the emission without lowering the S/N where lines are narrow. Otherwise, ${\sim}20\%$ of the sightlines detected in \hcnthree and/or \hcopthree would have been misclassified as non-detections. 

Specifically, we considered pixels where the line peak within the galaxy velocity range is higher than 1.5$\sigma_{\rm channel}$. We fitted a single Gaussian to the line in each pixel and afterwards: (1) If the fit simultaneously satisfied that its velocity centroid lied within the galaxy-wide velocity range, its FWHM did not exceed this range, and its area had a S/N $\geqslant 5$, then we placed the window limits at the 5\% level of the fitted Gaussian. (2) If any of these criteria were not fulfilled, we adopted the limits derived in the same way from the high-S/N \cotwo line. (3) If the \cotwo fit did not meet these criteria either, we adopted the galaxy velocity range. Once the velocity window was determined, we calculated the velocity-integrated moments of order 0 (integrated intensity, $I_\textrm{line}$), 1 (mean velocity), and 2 (velocity dispersion). Moment uncertainties follow the standard formulas of error propagation. 

We consider that a molecular line is detected when the S/N of the integrated intensity is ${\geq} 5$. This strict criterion mainly selects pixels with the velocity windows defined by method (1) and where methods (1) and (2) typically differ by ${\lesssim} 5$\%.

\subsubsection{Intensity-weighted cloud-scale \cotwo intensity}\label{sec_meanco}

At their $1.5''$--$1.7''$ (71-121~pc) resolution, the  \cotwo lines from PHANGS-ALMA are good proxies for cloud-scale properties \citep{sun18,sun20,ler21b}. This information is partly blurred by beam dilution if the lines are directly convolved to a coarser resolution. To remedy this problem, \citet{ler16} put forward an alternative scheme to construct intensity-weighted averages of cloud-scale properties from high resolution data. Here we follow the modified approach by \citet{gal18b} (their Equation~1) to construct the intensity-weighted average of the \cotwo integrated intensity over 440~pc apertures (hereafter $\langle I_\text{CO21}^{1''}\rangle_\mathrm{440pc}$ or just $\meanco$) from the 1.5$''$-1.7$''$-resolution maps. We correct $\meanco$ for galaxy inclination by applying a cosine factor, as in \citet{sun22}. Barring differences in the CO conversion factor, $\meanco$ would be a good proxy for the cloud-scale surface density, but also of the gas volume density when the spatial resolution is fixed \citep[e.g.][]{ler16,gal18b,uto18}.
 
\begin{figure*}[!ht]
\begin{tabular}{ccc}
\large\bf\ngc{3351} & \large\bf\ngc{3627} & \large\bf\ngc{4321}\\[0.3cm]
\multicolumn{3}{c}{\sc\large New \hcnthree and \hcopthree Maps}\\[0.2cm]
\includegraphics[width=0.33\linewidth]{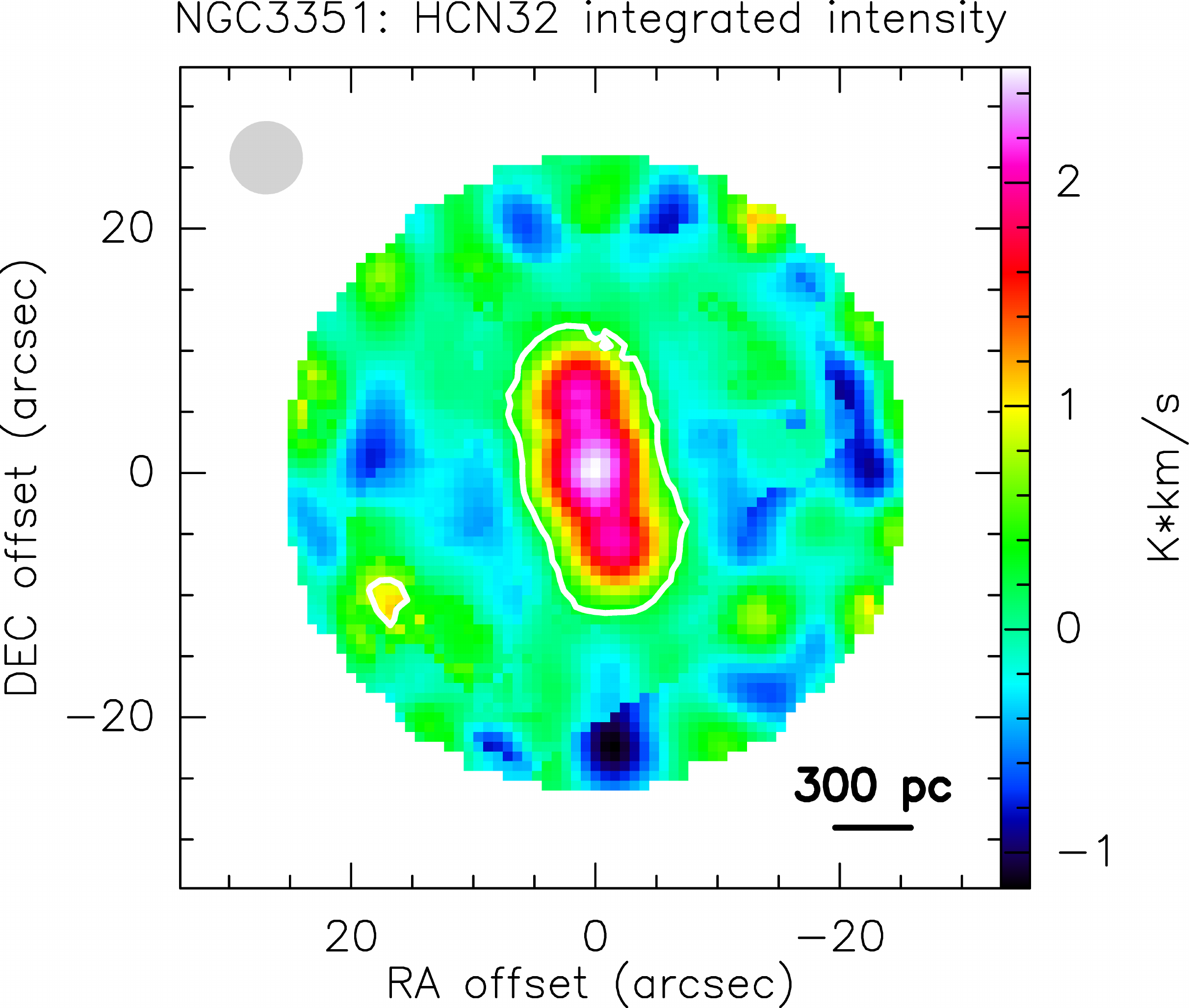} &
\includegraphics[width=0.27\linewidth]{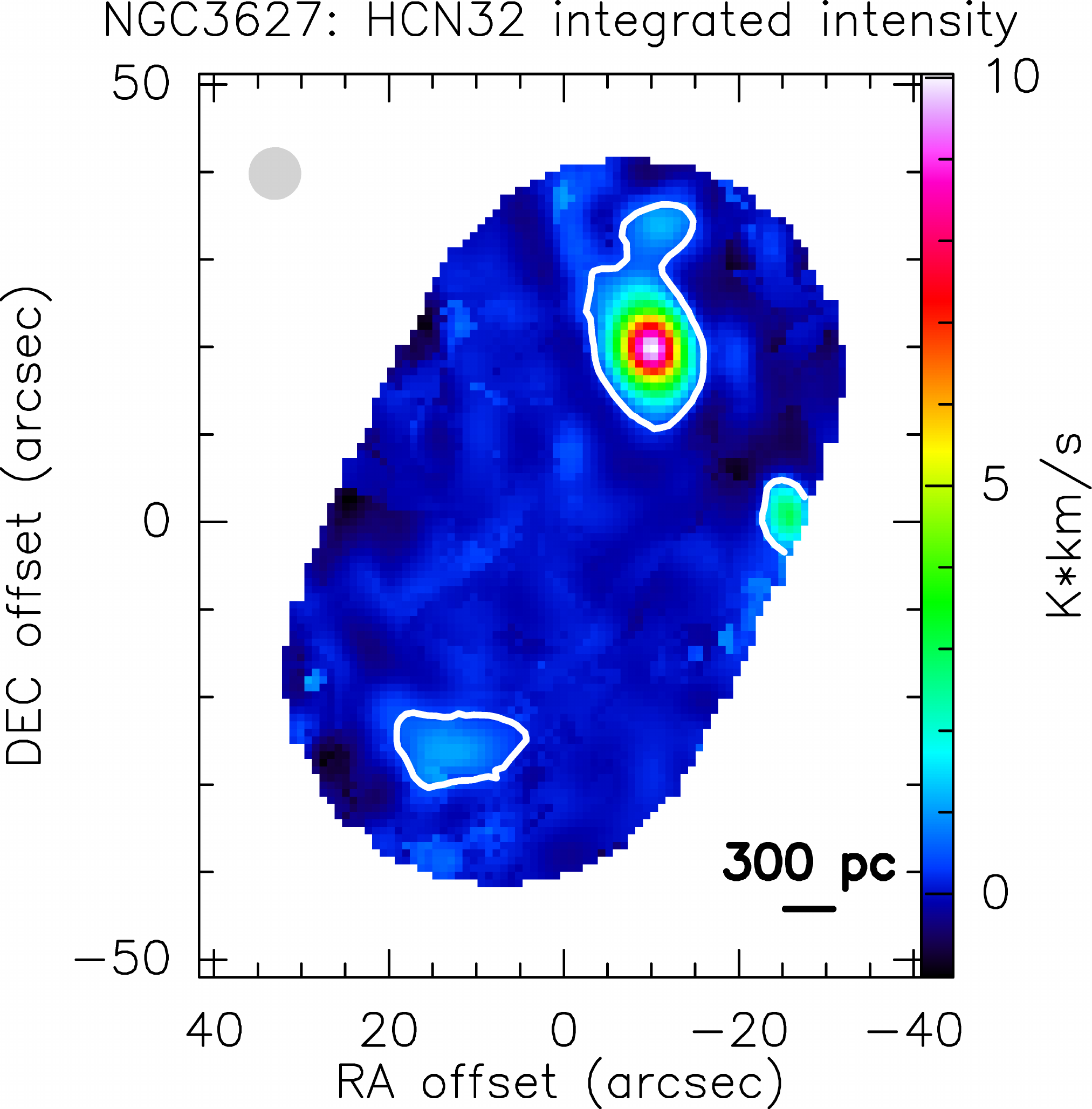} &
\includegraphics[width=0.33\linewidth]{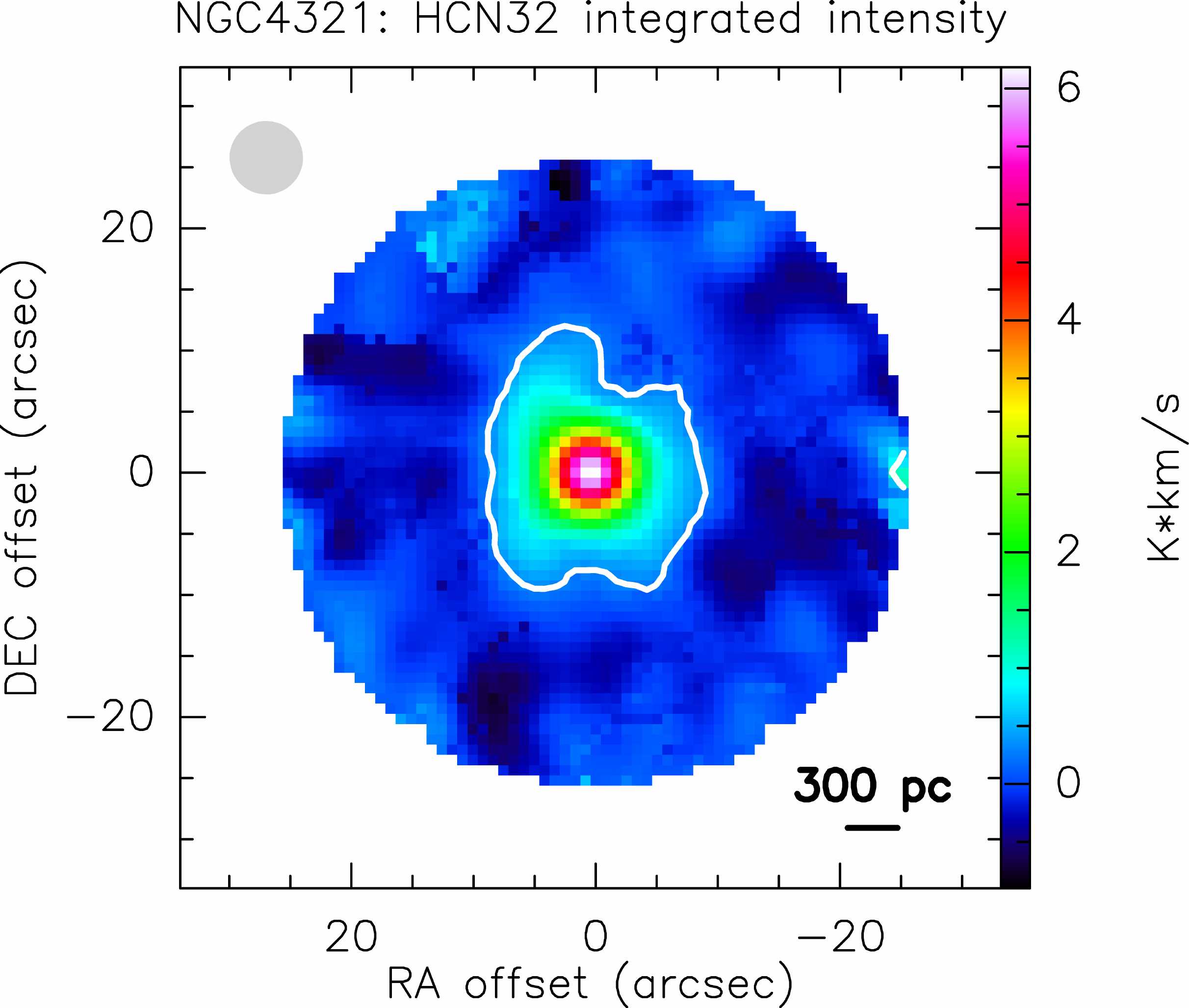} \\
\includegraphics[width=0.33\linewidth]{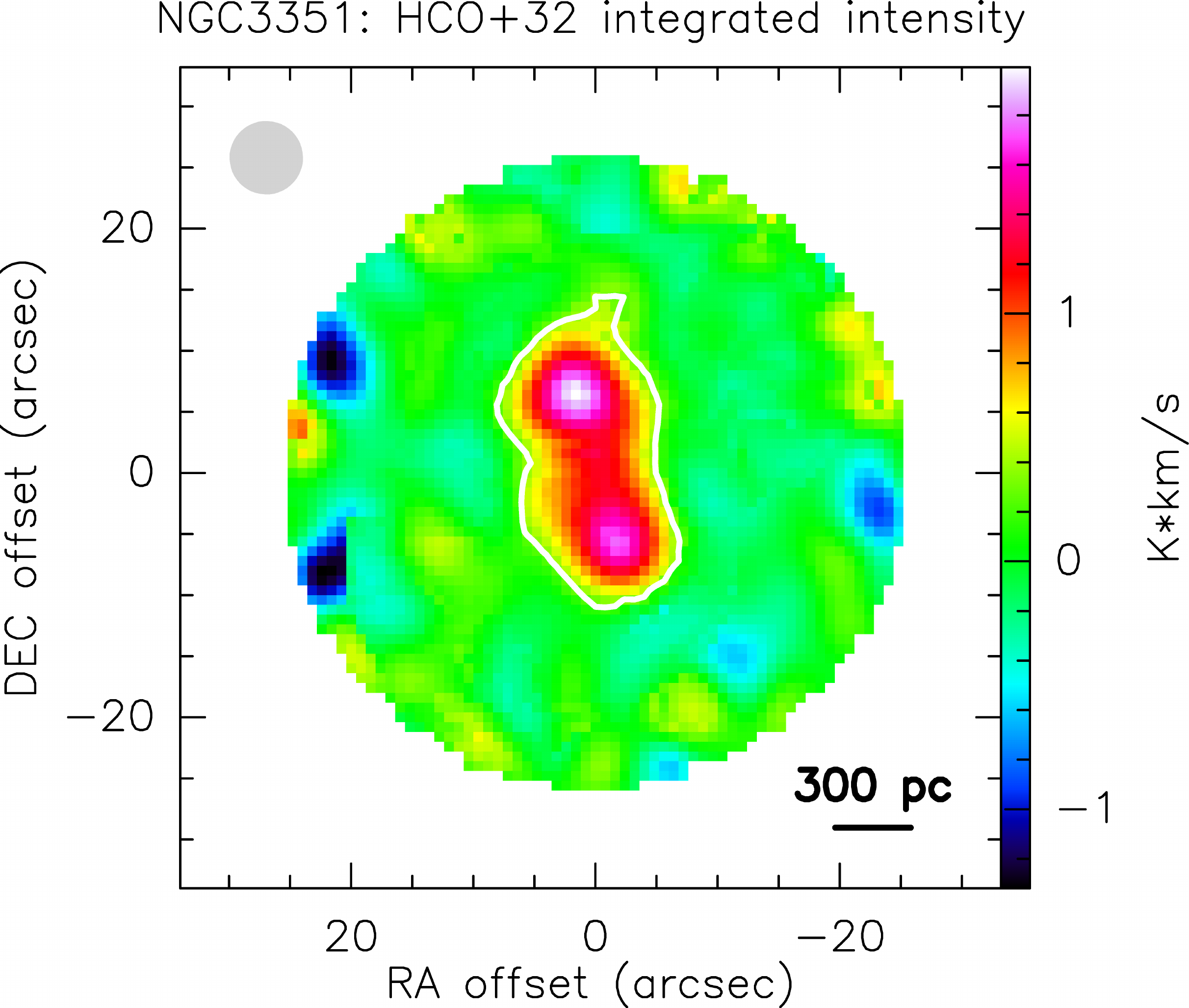} &
\includegraphics[width=0.27\linewidth]{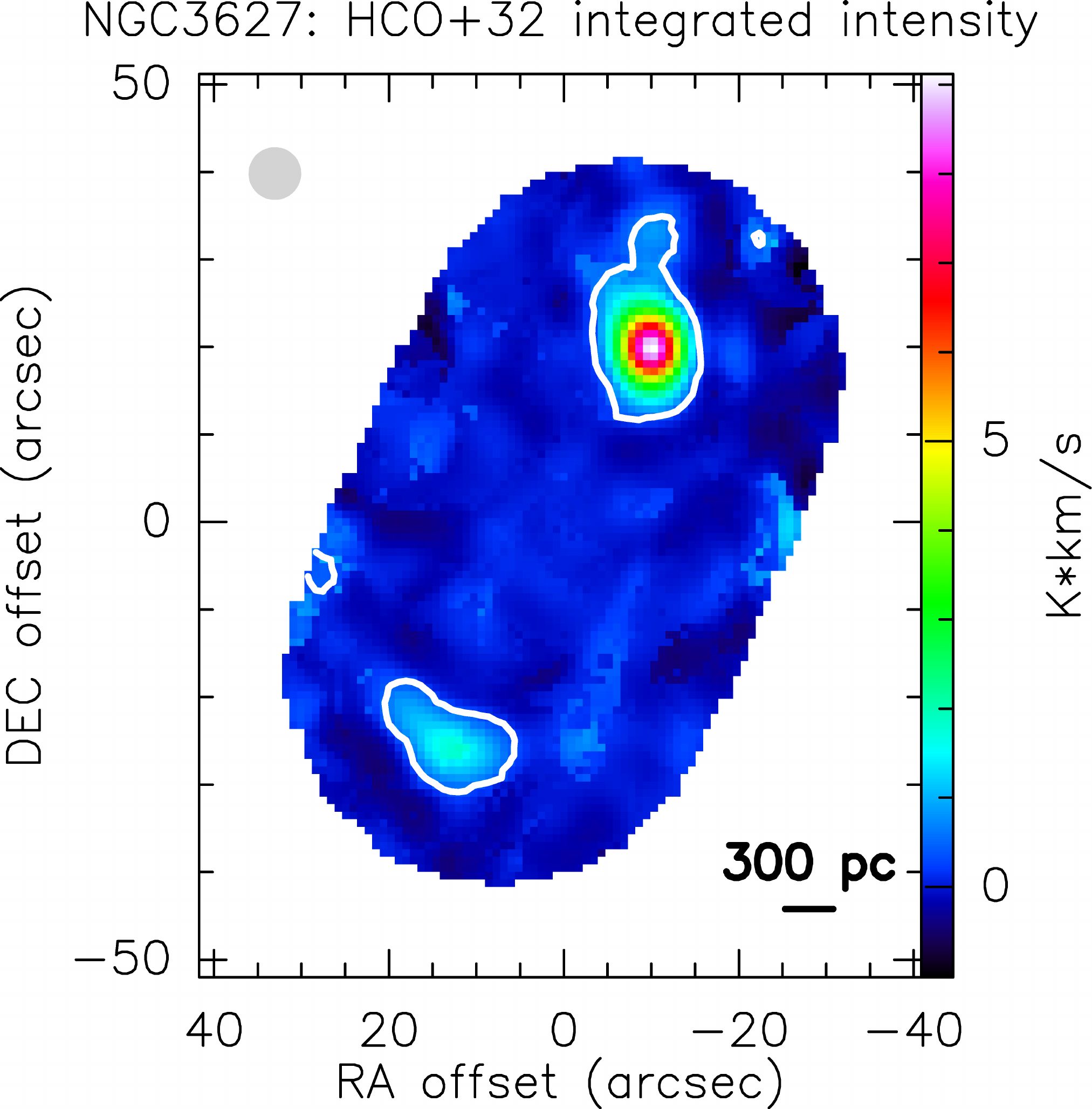} &
\includegraphics[width=0.33\linewidth]{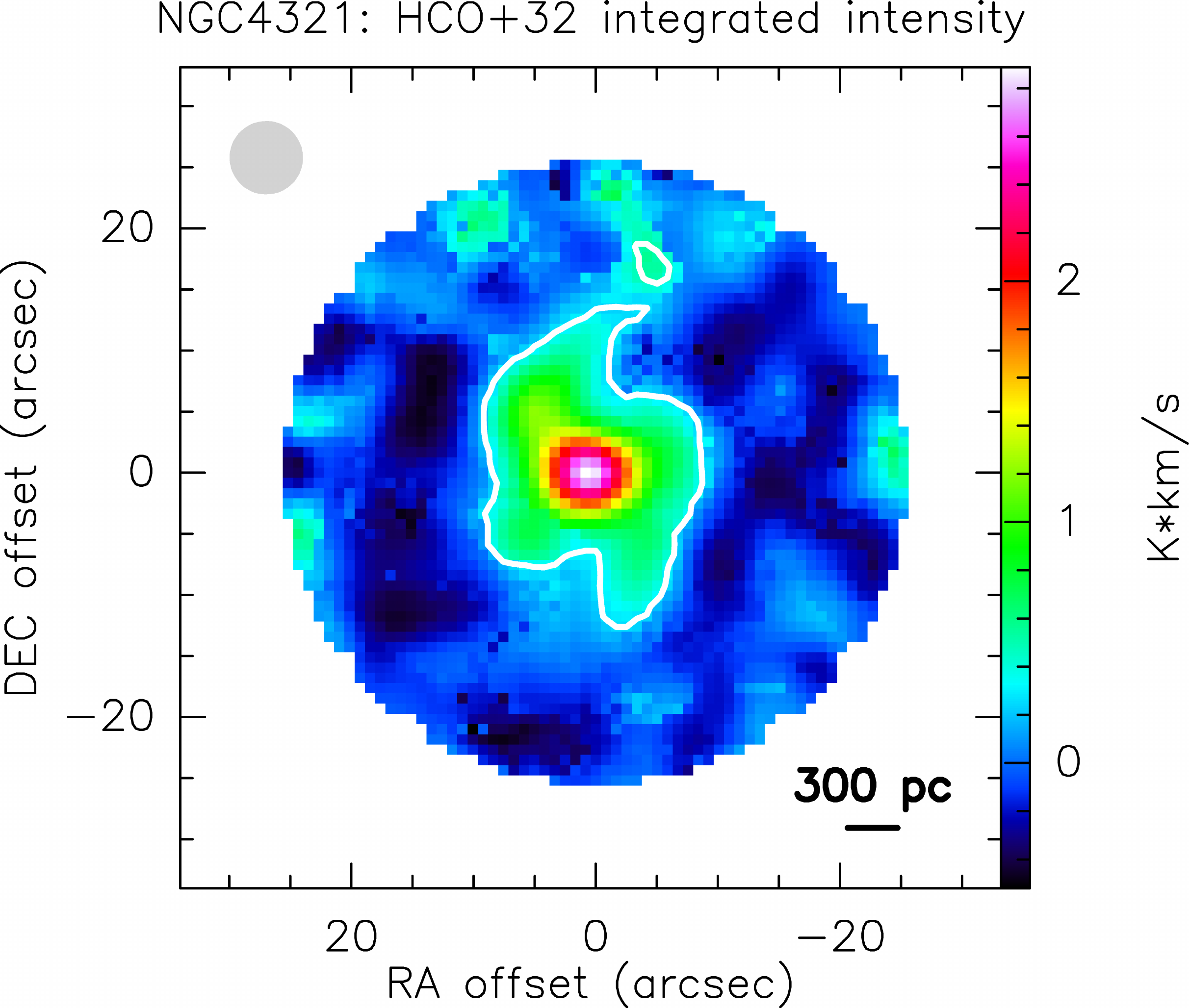} \\[0.3cm]
\multicolumn{3}{c}{\sc\large\cotwo Maps from PHANGS-ALMA}\\[0.2cm]
\includegraphics[width=0.32\linewidth]{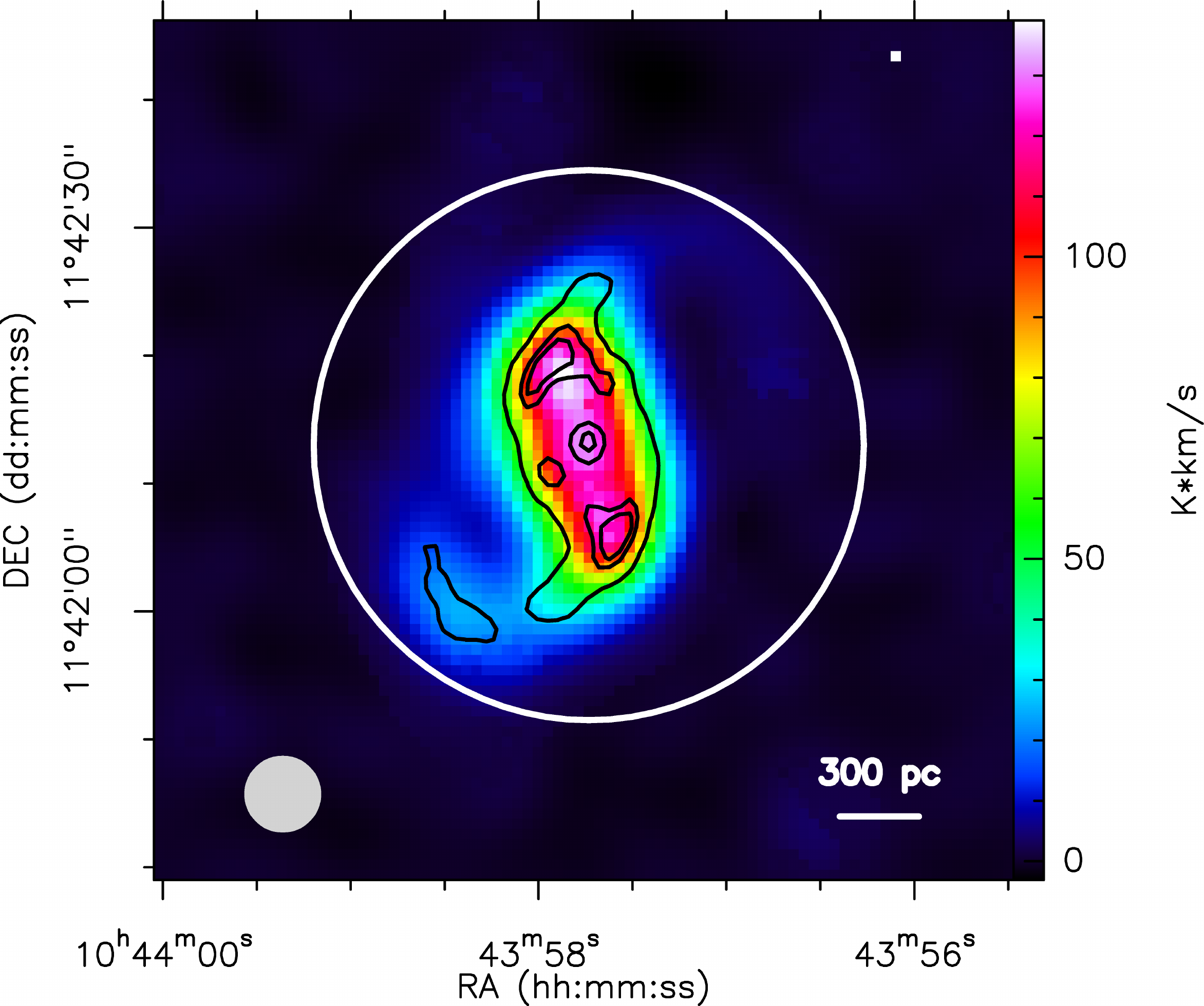} &
\includegraphics[width=0.26\linewidth]{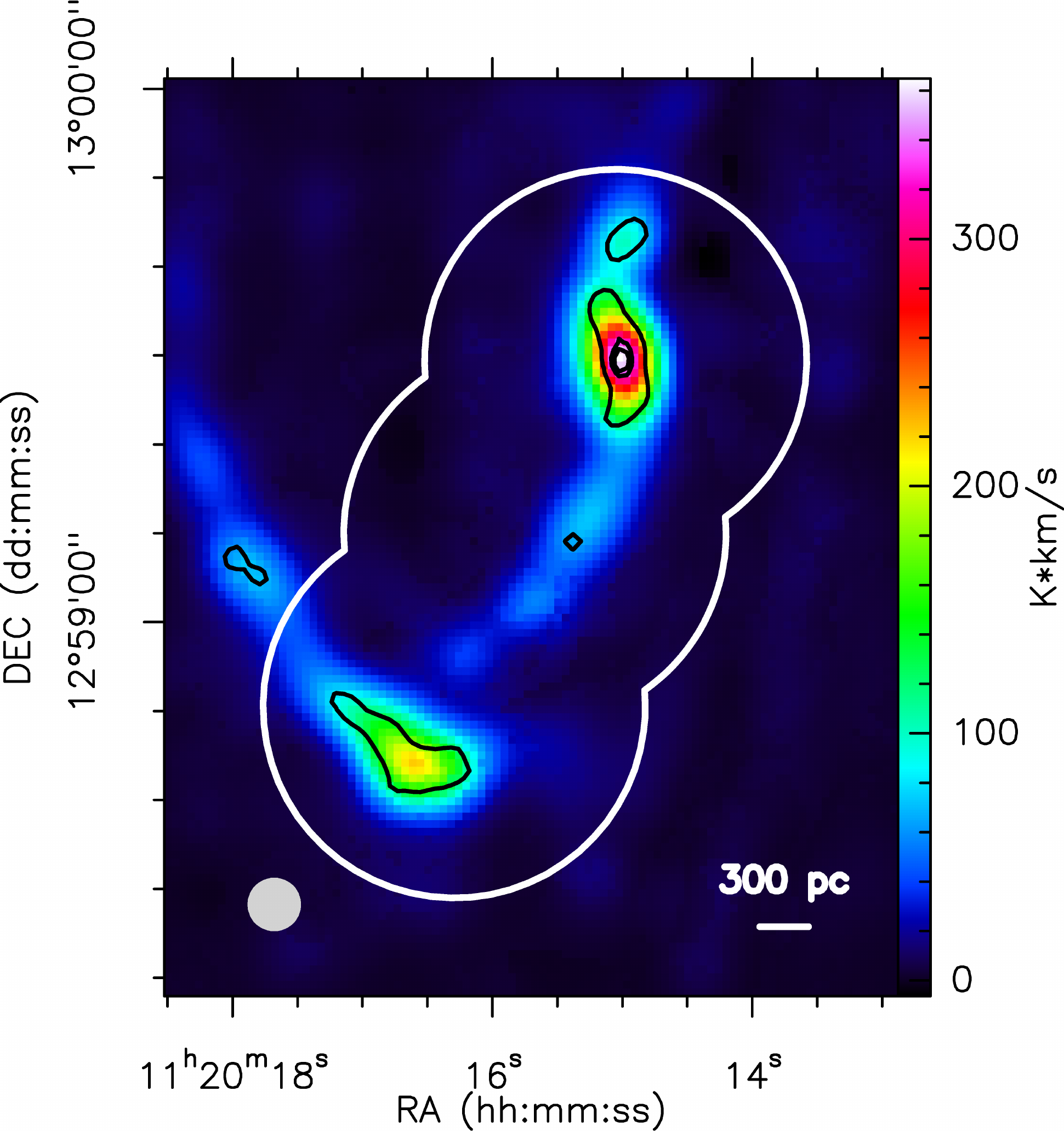} &
\includegraphics[width=0.32\linewidth]{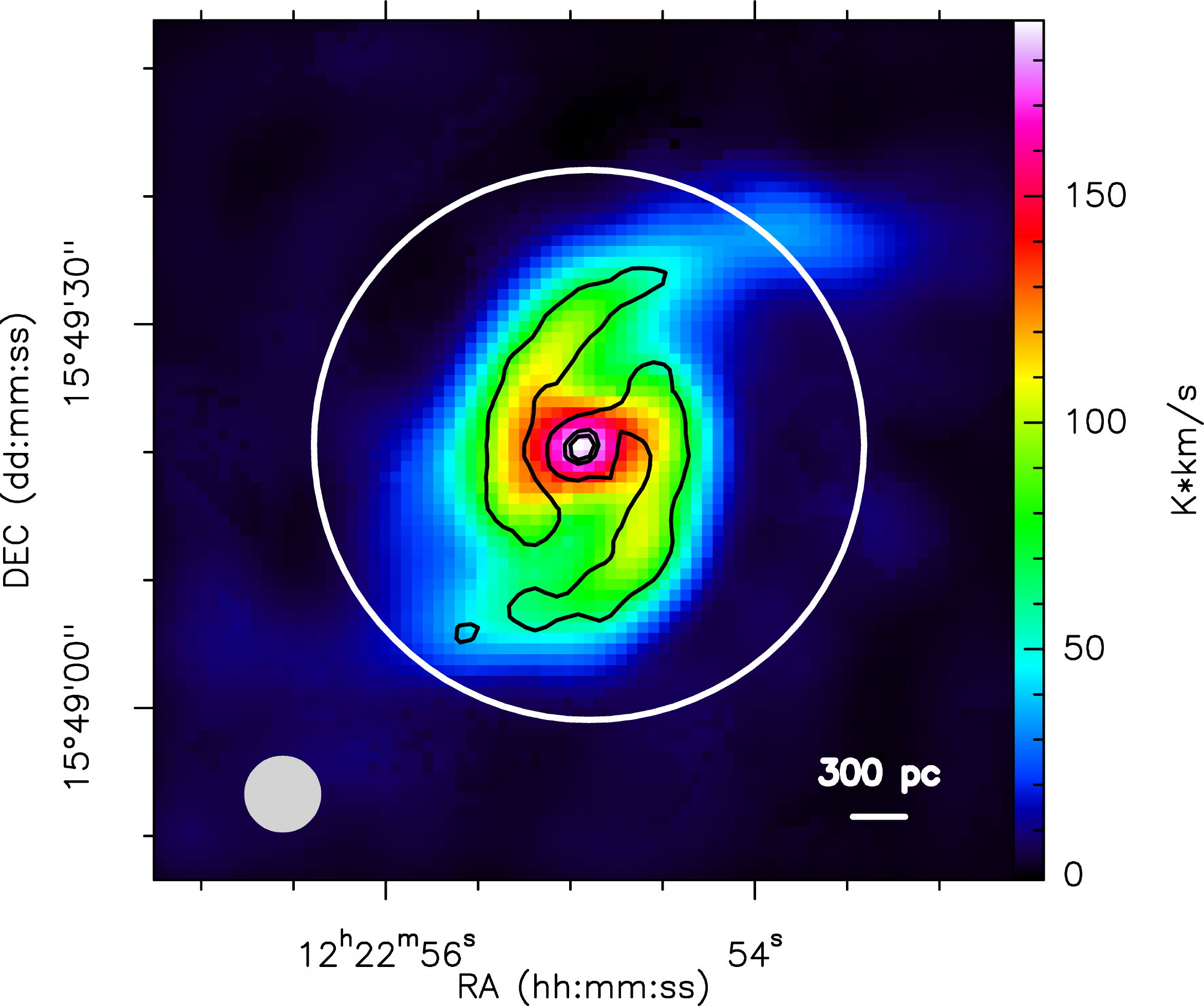} \\
\end{tabular}
\caption{
Integrated intensity maps at $6''$ resolution of \ngc{3351}, \ngc{3627}, and \ngc{4321 from left to right}. Top and middle rows: \hcnthree and \hcopthree  maps. The white contours correspond to S/N=5 and the grey circles represent the $6''$ beam. Offsets are relative to the positions in Table~\ref{tab_sample}. Bottom row: PHANGS-ALMA \cotwo maps. Black contours correspond to 10, 30, 50, and 70\% of the maximum in each \cotwo image at the native ${\sim}1.5''$ resolution. The white circles indicate the fields of the \hcnthree and \hcopthree observations with the ACA.}
\label{fig_mom0-HCN32}
\end{figure*}

\subsection{Ancillary environment and SFR data}\label{sec_ancillaryEnviron}

The \textbf{stellar surface density ($\sstar$)} traces the gravitational potential in the inner discs of massive galaxies, since stars dominate their mass budget \citep{cas17}. We derive $\sstar$ from the 3.6~$\mu$m continuum images taken with Spitzer/IRAC as part of the S$^4$G survey \citep{she10,mun13,que15}\footnote{\url{https://sha.ipac.caltech.edu/applications/Spitzer/SHA/}}. We subtracted a flat background equal to the mode value in signal-free regions. There was no need to remove any foreground star over the studied regions. The light-to-mass calibration follows \cite{ler08}: 
\begin{equation}\label{eq_sigmaStar}
\frac{\Sigma_\text{star}}{{M}_\odot \, \textrm{pc}^{-2}}= 280 \ \frac{{I}_{3.6\mu \textrm{m}}}{\textrm{MJy} \, \textrm{sr}^{-1}}\cos(i), 
\end{equation}
where $i$ stands for the disc inclination. This recipe disregards potential variations in the mass-to-light ratio and other contributions to the 3.6$\mu$m emission \citep{mei12,mei14}.

Our \textbf{star formation rate surface density ($\ssfr$)} combines unobscured (traced by  H$\alpha$) and obscured (traced by the 24$\mu$m continuum) emission from massive stars, as in \citet{ken09}. Other SFR tracers yield qualitatively the same results in our targets \citep{gal18}. We use the Convolved and OPTimised (COPT)  H$\alpha$ maps from the PHANGS-MUSE survey \citep[DR 2.2,][]{ems22}, which have a Gaussian point spread function (PSF) of ${\sim}1''$, and the 24~$\mu$m MIPS maps from the Spitzer archive, which we processed as the 3.6$\mu$m data. We combine their flux densities ($I$) to derive $\ssfr$, with H$\alpha$ typically accounting for $\lesssim$20\% of the total: 
\begin{equation}\label{eq_sfr}
\frac{\ssfr}{{M}_\odot \, \textrm{yr}^{-1} \, \textrm{kpc}^{-2}} = 945.43 \frac{{I}_{\text{H}\alpha}+0.02\,{I}_{24\mu \textrm{m}}}{\textrm{erg} \, \textrm{s}^{-1} \, \textrm{cm}^{-2} \, \textrm{sr}^{-1}}\cos(i).
\end{equation}

We convolved these data sets to  our $6''$ working resolution. We applied a Gaussian kernel to the H$\alpha$ images and the IDL routines by \citet{ani11} to the 3.6 $\mu$m and 24 $\mu$m maps.

\section{Integrated moments from the ACA band-6 data}\label{sec_rawResults} 

Figure~\ref{fig_mom0-HCN32} displays in colour scale the integrated intensity maps of the \hcnthree and \hcopthree lines. In the bottom row, the \cotwo maps and their corresponding high resolution contours (${\sim}1.5''$) are shown as reference. Their higher S/N and spatial resolution help us identify the morphological features described in Sect.~\ref{sec_sample}. 
We detect and resolve the \hcnthree and \hcopthree emission at S/N$\gtrsim5$ (white contours) over: (1) the inner star-forming ring of \ngc{3351}; (2) the centre and the southern bar end of \ngc{3627}; and (3) the centre and the base of the spiral arms of \ngc{4321}. To first order, the spatial distributions of the \hcnthree and \hcopthree lines are similar to that of \cotwo, though restricted to more compact regions. The latter results from a combination of lower S/N and steeper radial gradients in the integrated intensities (Sect.~\ref{sec_empirical-relations}). 
In closer detail, Figure~\ref{fig_mom0-HCN32} also shows some significant differences between \hcnthree and \mbox{\hcopthree} that we discuss in Sects.~\ref{sec_lineRatios} and \ref{sec_empirical-relations}.

\begin{figure}[!ht]
\centering
\includegraphics[width=\linewidth]{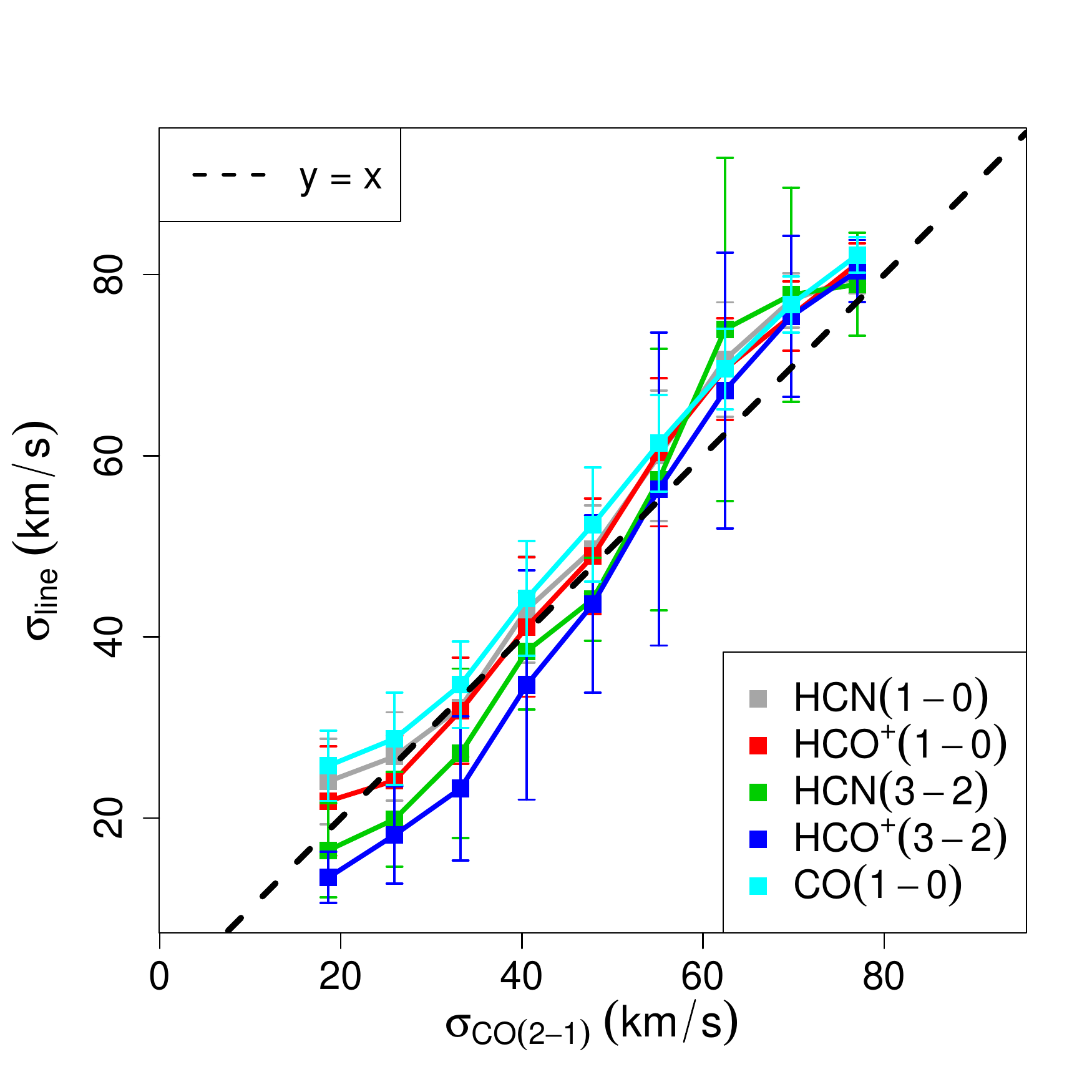} 
\caption{Velocity dispersion of each observed line compared with that of \cotwo at $6''$ resolution. We split the data into ten bins and show the means (squares) and the $\pm\sigma$ standard deviations (bars) for each bin. The black dashed line represents the 1-to-1 relation.}\label{fig_mom2-vs-co21}
\end{figure}

We do not present mean velocity nor velocity dispersion maps. However, we have inspected them to confirm that all lines listed in Table~\ref{tab_lineCatalogue} have similar spectral shapes on a pixel-by-pixel basis. The mean velocities of the different lines are in good agreement in the three targets. In the worst case, each line shows a median deviation in absolute value of $\sim$16~km~s$^{-1}$ with respect to \mbox{\cotwo}. This amounts to only 1.6 times the channel width. We check the consistency of the velocity dispersion of the lines in  \figcita{fig_mom2-vs-co21}. Specifically, we bin the sightlines of all three galaxies by the velocity dispersion of the high-S/N \cotwo line and represent the mean and the standard deviation for each line within each bin. Here, the lowest values ($\sim$15--20~km~s$^{-1}$) are typical of the discs, while the highest ones $\sim$50--100~km~s$^{-1}$) are found towards centres, where the unresolved rotation curve gradients and/or the increased turbulence broaden the linewidths \citep{she12,sun20, miu21, kri21}. We find that all trends are mutually consistent within the scatter, since they agree within $\lesssim$25\% with the one-to-one relation (with some systematic deviations both at lower and higher values). In summary, we can reasonably consider that all lines have similar shapes, so that line ratios do not strongly depend on velocity. Thus, we can safely study the excitation conditions along each sightline from the integrated intensity maps.

\section{Line ratios as a probe of the dense gas excitation}\label{sec_lineRatios}

We construct line ratio maps from the integrated intensity maps of the six lines in Table~\ref{tab_lineCatalogue}. We use them to explore how the excitation conditions vary across our targets. Specifically, we focus on: (1) Same-species ratios (e.g. \hcnthree/\hcnone, \hcopthree/\hcopone), which represent our most direct proxies for the excitation of each molecule; (2) Same-transition ratios (e.g. \hcnthree/\hcopthree, \hcnone/\hcopone), which help us single out the effects of chemistry; (3) HCN/CO ratios (e.g. \hcnthree/\cotwo, \hcnone/\cotwo), which are frequently used to infer dense gas fractions in external galaxies (Sect.~\ref{sec_intro}). Other possible ratios are combinations of these ones or lack an intuitive interpretation.

Line ratios are sensitive to several parameters such as density, chemical abundances, opacity or temperature. Here, we define the line ratios so that the critical density is higher for the numerator than for the denominator (Table~\ref{tab_lineCatalogue}), which helps us discuss the potential effects of density. Unless otherwise stated, in the following we only take lines of sight (pixels) where all lines are detected at $\geq5\sigma$ significance into account. This criterion suppresses possible noise-driven artefacts, although it can introduce a selection bias that we discuss whenever appropriate.

\begin{figure*}[ht!]
\centering
\includegraphics[width=0.9\linewidth]{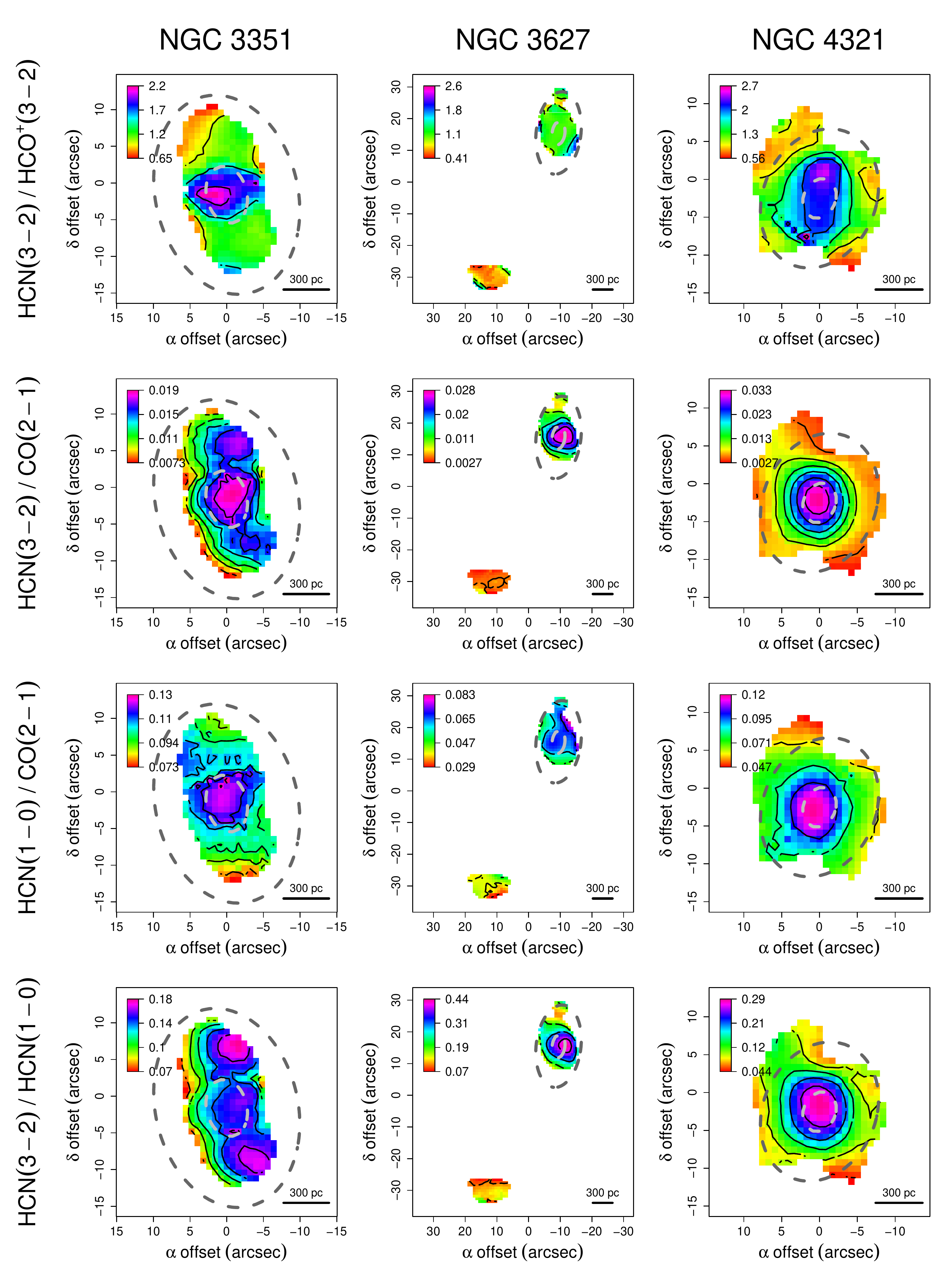} 
\caption{
Colour maps of the \hcnthree/\hcopthree, \hcnthree/\cotwo, \hcnone/\cotwo, and \hcnthree/\hcnone line ratios from top to bottom. We only represent pixels where all lines are detected at $\geq$5$\sigma$. The grey dashed ellipses indicate galactocentric radii of 200 and 700 pc. Black contours correspond to five equispaced levels from 17\% to 83\% of the maximum value in each panel. }\label{fig_lineRatioMaps}
\end{figure*}

Figure~\ref{fig_lineRatioMaps} displays the maps of a subset\footnote{Here we skip the \hcopthree/\hcopone and \hcnone/\hcopone maps. As in Sect.~\ref{sec_litComp}, they are similar to the \hcnthree/\hcnone and \hcnthree/\hcopthree ones, respectively.} of the line ratios that we discuss in Sects.~\ref{sec_lineRatios}-\ref{sec_empirical-relations}. To first order, these ratios peak at the galaxy centres and tend to decrease outwards, although we recognise in most panels the  morphological features that were apparent in Fig.~\ref{fig_mom0-HCN32}. The only exception to this radial behaviour is the \hcnthree/\hcnone ratio in \ngc{3351}, which actually peaks on the two contact points of the star-forming ring.

\subsection{HCN and HCO$^+$ line ratio statistics}\label{sec_litComp}

\begin{figure*}[!ht]
\includegraphics[trim= 0 0 0 0, clip,width=\linewidth]{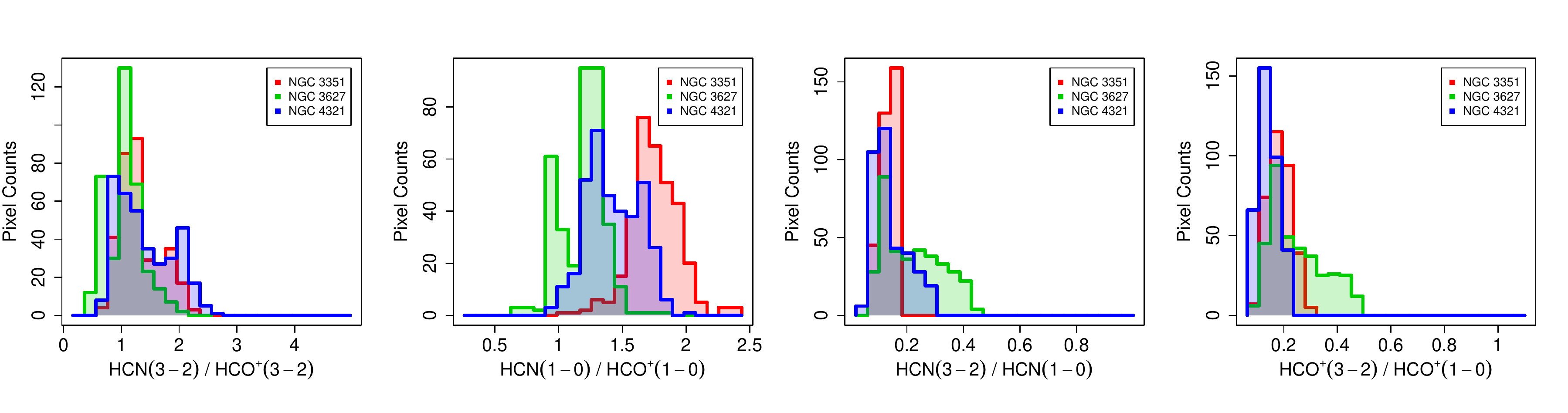}
\caption{Histograms per galaxy of \hcnthree/\hcopthree, \hcnone/\hcopone, \hcnthree/\hcnone, and \hcopthree/\hcopone line ratios.}\label{fig_litHist-simple}
\end{figure*}

\begin{figure*}[!ht]
\includegraphics[trim= 0 14cm 0 0, clip,width=\linewidth]{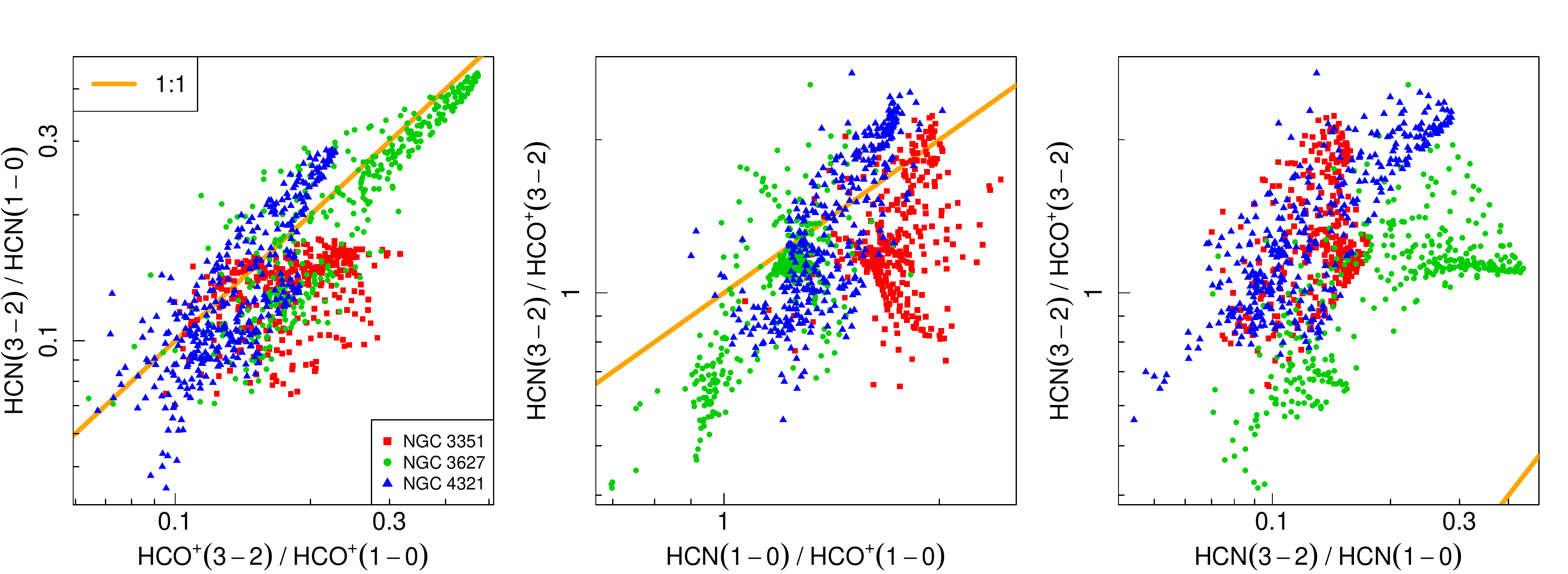} 
\caption{Ratio vs ratio diagnostic plots combining the HCN and HCO$^+$ $J$=1--0 and 3--2 lines. Each dot corresponds to one pixel in our galaxies, coloured by target. The orange lines are one-to-one linear relations.}\label{fig_dp}
\end{figure*}

Figure~\ref{fig_litHist-simple} presents the histograms per galaxy of line ratios \mbox{\hcnthree/\hcopthree}, \hcnone/\hcopone, \mbox{\hcnthree/\hcnone}, and \hcopthree/\hcopone (the statistics of these and other ratios is also summarised in Appendix~\ref{app_SummlineRatios} as a reference for future studies). The plots show that the maximum span of each ratio over our  sample is $\sim$0.6-1~dex. Part of the broad scatter comes from variations within each galaxy, since the multiple peaks or wings of the histograms are associated with different morphological features (Fig.~\ref{fig_lineRatioMaps}). In addition, some ratios exhibit significant galaxy-to-galaxy differences. Clear examples are the almost disjoint histograms of \hcnone/\hcopone in \ngc{3627} and \ngc{3351}. Figure~\ref{fig_dp} presents a complementary view, where the ratios are plotted against each other. The leftmost panel shows that, to first order, the two \mbox{3--2/1--0} ratios  tend to align with the same 1-1 relation. This implies that the excitation of HCN and HCO$^+$ is covariant to a significant degree, although we still see a non-negligible scatter in our data. We can infer similar conclusions from the HCN-to-HCO$^+$ ratios shown in the middle panel. In contrast, the rightmost panel shows a weaker correlation between \hcnthree/\hcopthree and \hcnthree/\hcnone, which actually might be induced by \hcnthree being present in the two axes.

Our ranges of fixed-transition HCN/HCO$^+$ ratios are comparable to those reported in the literature, dominated by the active galactic nuclei (AGN) and starburst galaxies, whereas our fixed-species 3--2/1--0 ratios tend to be lower. However, these comparisons could be biased by differences in resolution and the lack of complete studies (Sect.~\ref{sec_intro}). Larger and more homogeneous samples are needed to robustly assess any differences in excitation. Figure~\ref{fig_litHist} is an expanded version of Fig.~\ref{fig_litHist-simple} that includes data from a comprehensive set of studies.

\subsection{Comparison with non-local thermodynamic equilibrium models}
\label{sec_radex}

We run basic non-local thermodynamic equilibrium (non-LTE) calculations to explore the excitation regime of HCN and HCO$^+$ in our targets. We use RADEX \citep{tak07} for radiative-transfer calculations, supplied with radiative and collisional coefficients from the Leiden Atomic Molecular DAtabase \cite[LAMDA;][]{sch05}. Escape probabilities assume an homogeneous sphere geometry. We disregard the hyperfine splitting of the HCN rotational lines and only consider collisions with H$_2$ molecules. For HCN, we build a grid of single-phase RADEX models that spans a broad range of each of the three free parameters: the kinetic temperature ($\tkin$; 10--100~K), the H$_2$ volume density ($n$; 10--10$^8$~cm$^{-3}$), and the HCN abundance per velocity gradient ([HCN]/$\vgrad$;  10$^{-10}$--10$^{-6}$~pc~(km~s$^{-1}$)$^{-1}$). We build a parallel HCO$^+$ grid on the same assumptions.

Out of the grid of HCN models, we select those consistent with the observations in our targets. Specifically, we choose models that predict: (1) \hcnthree/\hcnone ratios consistent with the 5$^\mathrm{th}$--95$^\mathrm{th}$ percentile range of the observations (0.08--0.35); (2) \hcnthree brightness temperatures $\geq6$~mK (our 3$\sigma$ level)\footnote{Dilution in area and velocity could make the observed brightness temperature lower than the intrinsic one, but not higher.}. The selected models invariably predict \hcnthree lines that are subthermally excited and, in a clear majority of cases, optically thick. The predicted \hcnone lines tend to be optically thicker and closer to thermalisation\footnote{i.e. the $J_\nu(T_\mathrm{ex}) - J_\nu(2.73\,\mathrm{K})$ factor is closer to its thermal value ($T_\mathrm{ex}$=$\tkin$) for \hcnone than for \hcnthree. $J_\nu$ and $T_\mathrm{ex}$  are the radiative transfer equation and the excitation temperature of each transition, respectively.}. We filtered the HCO$^+$ grid in a similar manner and found similar conclusions.

This simple comparison provides us with some indications of the dependence of the lines on the excitation parameters. For example, in the regime of subthermal optically thick emission, HCN brightness temperatures are sensitive to $\tkin$, but also to the number of photons generated by collisions, which tends to enhance the excitation temperature and is proportional to $n^2[\mathrm{HCN}]/\vgrad$ \citep{sco74}. The dependence on the latter parameter weakens as the lines approach thermalisation, so it tends to be stronger for \hcnthree than for \hcnone.

\section{Empirical line ratio relations}\label{sec_empirical-relations}

Observations of \hcnone and \hcopone in nearby galaxies at (sub-)kiloparsec scales have revealed systematic variations in the line/CO and $\ssfr$/line ratios as a function of environment parameters such as $\sstar$ \citep[e.g.][]{use15,big16,gal18,jim19,que19,san22b}, or cloud-scale properties such as $\meanco$ \citep{gal18b,neu22}. These relations have important implications for the internal and SF properties of the clouds, on the assumption that the \hcnone and \hcopone lines are reliable proxies for the dense gas mass (Sect.~\ref{sec_intro}). In this section, we compare these relations with those inferred from  our \hcnthree and \hcopthree observations. Only for these purposes, we have homogenised the data of the three galaxies to their best common physical resolution (440~pc $\approx$ 6$''$--9$''$).

\begin{figure*}[!ht]
\centering
{\sc Empirical gas-environment relations at Fixed 440 pc Resolution}\\[-0.3cm]
\includegraphics[width=.8\linewidth]{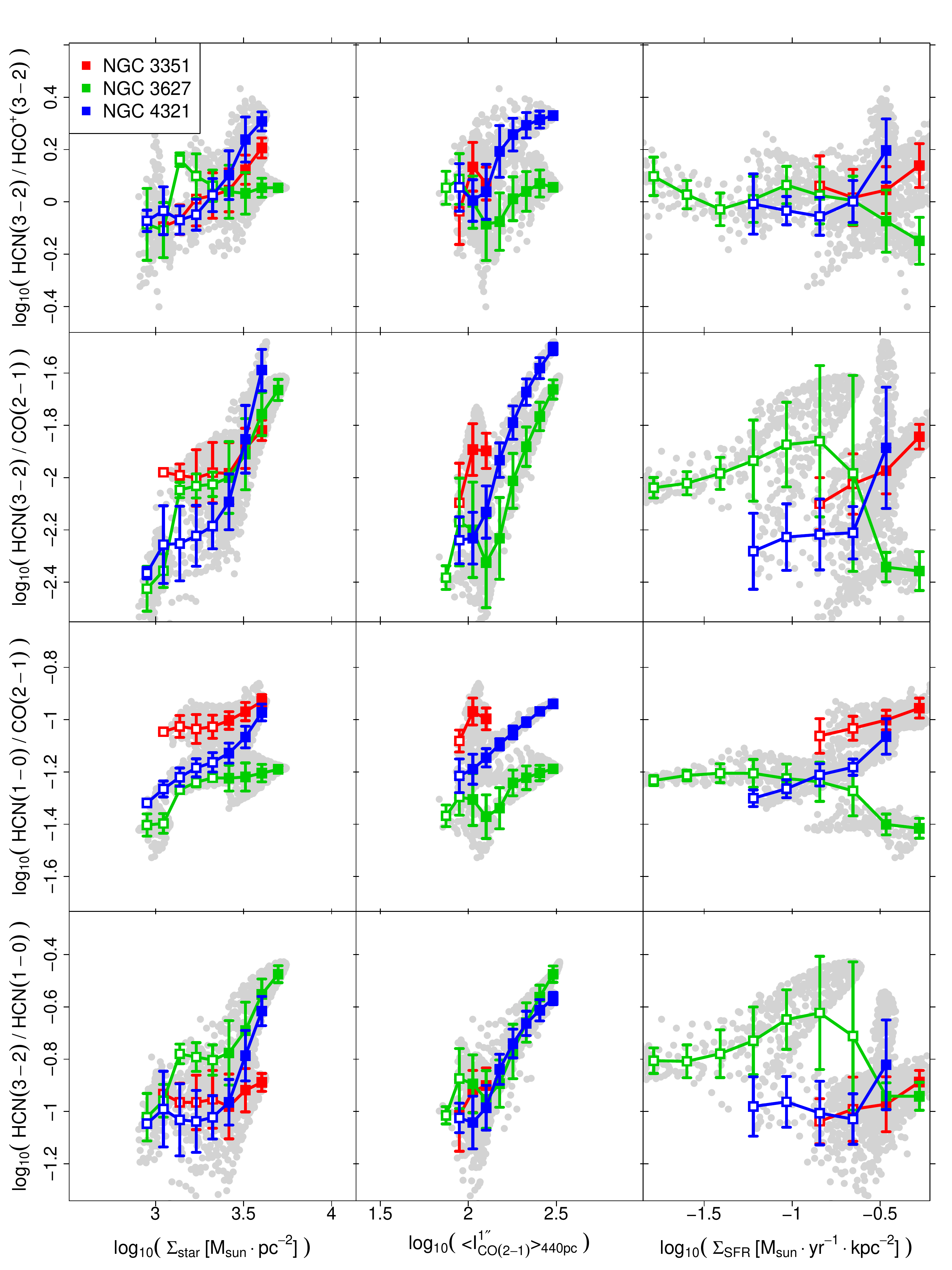} 
\caption{Same line ratios as in Fig.~\ref{fig_lineRatioMaps} plotted against $\sstar$ (left), $\meanco$ (middle), and $\ssfr$ (right) on a pixel-by-pixel basis (grey points). 
All panels have the same vertical and horizontal span in dex for an easier comparison. 
For each line ratio and galaxy, the colour squares and vertical bars represent the  mean  and the $\pm 1\sigma$ standard deviation within bins of the x-axis parameter. The squares are white-filled where the mean value does not necessarily represent the whole galaxy (Sect.~\ref{sec_empirical-relations}).}\label{fig_line-ratio_galaxy_comparison}
\end{figure*}

Figure~\ref{fig_line-ratio_galaxy_comparison} shows the systematic variations over the three targets in: the $J$=3--2 HCN/HCO$^+$ ratio, the two HCN/CO line ratios, and the HCN 3--2/1--0 excitation ratio. We plot them as a function of $\sstar$ (left), $\meanco$ (middle), and $\ssfr$ (right). Once differences in $\sstar$ calibration or CO transition are taken into account, the \hcnone/CO relations are consistent with \citet{gal18b,gal18}, where these data were taken from, as well as with other fits derived from larger galaxy samples \citep[e.g.][]{use15,jim19}.

To discuss how line ratios vary, we focus on the binned trends shown in the figure. They represent the overall behaviour over the S/N-selected pixels inside the ACA FoV, but  do not necessarily represent the behaviour over an entire galaxy. To mark where this might happen, we proceed as follows. For each $\sstar$ bin (or other x-axis quantity) we sum the total \cotwo flux from all the pixels that belong in the bin and lie: either (1) anywhere in the galaxy, or (2) only in the region that we selected for analysis. When the (2)-to-(1) ratio is less than 0.8, we consider that the bin value is not necessarily representative, which we indicate with a white-filled square. In some panels the trends bend where white-filled squares set in, suggesting that these values suffer from a selection bias.

Within the limited dynamic range covered by the data, most panels of Figure~\ref{fig_line-ratio_galaxy_comparison} (left and middle) show that the line ratios are enhanced when $\sstar$ and/or $\meanco$ increase. The overall agreement between $\sstar$ and $\meanco$ trends mostly reflects that the stars and CO emission are similarly distributed over galaxy discs \citep[e.g.][]{ler08}. The main differences between rows are the steepness of the trends in individual galaxies and the importance of galaxy-to-galaxy offsets. On the one hand, there is a range of $\sim$2 in the overall slopes, with the sense that \hcnthree/\cotwo > \hcnthree/\hcnone > \hcnthree/\hcopthree $\approx$ \hcnone/\cotwo. On the other hand, galaxy-to-galaxy offsets up to $\sim$0.3~dex are apparent in some panels and can contribute significantly to the overall scatter. The clearest exception is the \hcnthree/\hcnone--$\meanco$ relation shown in the bottom-middle panel, where the three galaxies are well aligned. This tight correlation is an example that line ratios can track the average properties of the molecular gas measured at higher spatial resolution \citep{gal18b}.

Unlike in the $\sstar$ and $\meanco$ panels, there are no robust trends in the $\ssfr$ panels, which also show a higher scatter per bin. This suggests that the energy input provided by star formation does not have a net impact on the excitation of the lines (Sect.~\ref{sec_radexDiscussion}). We cannot discard that mechanisms other than SF might operate or even dominate gas heating in our regions of study \cite[e.g. turbulence dissipation, cosmic rays;][for examples in the Galactic CMZ]{mil17}.

\section{Discussion: Drivers of line ratio relations}\label{sec_radexDiscussion}

We briefly discuss some implications of Fig.~\ref{fig_line-ratio_galaxy_comparison} for the excitation of HCN (they apply to HCO$^+$ as well). We focus on the trends against $\meanco$, which are barely affected by the potential selection biases mentioned in Sect.~\ref{sec_empirical-relations} and are easier to interpret in terms of cloud properties. $\meanco$ is a good proxy for density (Sect.~\ref{sec_meanco}) and, since \cotwo is optically thick and easy to thermalise under normal conditions, it is also sensitive to $\tkin$. For the sake of the argument, let us assume that all trends in each galaxy were driven by density, with other excitation parameters kept fixed. Based on Sect.~\ref{sec_radex}, the different thermalisation degree of \cotwo (highest), \hcnone, and \hcnthree (lowest) would qualitatively explain why \hcnthree/\cotwo, \hcnone/\cotwo, and \hcnthree/\hcnone tend to increase  with $\meanco$, and why the \hcnthree/\cotwo trends are steeper than the \hcnone/\cotwo ones. Within this picture, differences in $\tkin$ between galaxies could shift  $\meanco$, thus generating galaxy-to-galaxy offsets. However, those offsets should  then be similar for all ratios, which is not the case (e.g. \hcnthree/\hcnone versus \hcnthree/\cotwo). Therefore, line ratios should also be sensitive to the parameter driving the offsets ($\tkin$ in this example).

This particular example illustrates that, at a more general level, the line ratios in Fig.~\ref{fig_line-ratio_galaxy_comparison} must be regulated by at least two parameters. Density and temperature are two appealing candidates, since both directly can affect the x- and y-axis quantities. To quantify their relative importance we must model the line ratios on a pixel-by-pixel basis, which motivates our Paper~II.

\section{Summary}\label{sec_Conclusions}
We present new ACA observations of the \hcnthree and \hcopthree line emission in the nearby SFGs \ngc{3351}, \ngc{3627}, and \ngc{4321}. At a spatial resolution of $\sim$290--440~pc, we detected and resolved the two lines in the inner $R_\textrm{gal} \lesssim$1~kpc of the three targets, as well as in the southern bar end of \ngc{3627}. We complemented these data with available interferometer images of the \coone, \cotwo, \hcnone, and \hcopone lines. This data set enabled an analysis of the empirical relations of HCN and HCO$^+$ in SFGs at sub-kiloparsec scales. This kind of analysis has been typically focussed on brighter AGN and starburst galaxies; in spite of that, SFGs host most of the SF in the local Universe. We focus on the behaviour of the set of line ratios (\hcnthree/\hcopthree, \hcnthree/\cotwo, \hcnone/\cotwo, and \hcnthree/\hcnone). We find the following main results:
\begin{itemize}
    \item The chosen line ratios peak at the galaxy centres and tend to decrease mildly as we move outwards. The only exception is \hcnthree/\hcnone in  \ngc{3351}, which rather peaks at two off-centre contact points between the  inner ring and the dust lanes of the bar ($R_\textrm{gal}\simeq$400~pc).
    \item For analysis, we selected pixels where all lines are detected at a $\geq$5$\sigma$ significance. Over them, the ratios vary by $\lesssim$0.6--1~dex (total range), with median values of 0.14 for \hcnthree/\hcnone, 0.17 for \hcopthree/\hcopone, 1.17 for \hcnthree/\hcopthree, and 1.40 for \hcnone/\hcopone.
    \item We compared our observations with a grid of one-zone non-LTE radiative transfer models from RADEX. We varied the kinetic temperature (10--100~K), the H$_2$ volume density (10--10$^8$~cm$^{-3}$), and the HCN abundance per velocity gradient (10$^{-10}$--10$^{-6}$~pc~(km~s$^{-1}$)$^{-1}$). The models compatible with our HCN observations predict that \hcnone and \hcnthree are subthermally excited and likely optically thick, with the former being somewhat closer to thermalisation.
    \item We explored the mean trends of our reference line ratios as a function of $\sstar$ (representing the environment), $\meanco$ (cloud-scale properties) and $\ssfr$ (SF) in each galaxy. No line ratio correlates with $\ssfr$, which implies that either they are not significantly affected by temperature or that SF does not dominate the gas heating in our target galaxies. In contrast, most ratios increase as a function of both $\meanco$ and $\sstar$ in each galaxy, with the former being less affected by selection biases and, thus, more robust.
    \item Within each galaxy, the slopes of the trends (in the log-log space) vary by a factor of $\sim$2, with the \hcnthree/\cotwo-$\meanco$ relations being two times steeper than \hcnone/\cotwo-$\meanco$. The trends also show $\lesssim$0.3~dex galaxy-to-galaxy offsets that can dominate the total scatter in some cases (e.g. \hcnone/\cotwo-$\meanco$). An exception is \hcnthree/\cotwo-$\meanco$, for which the offsets are negligible.
\end{itemize}

We finally discuss the implications of the different slopes and galaxy-to-galaxy offsets of these empirical relations. We conclude that the systematic variations in the HCN/CO and \hcnthree/\hcnone line ratios are not strictly ruled by a single excitation parameter. However, a detailed modelling is needed to separate the effects of parameters such as density and temperature in each case.

\begin{acknowledgements}
We are very thankful to the anonymous referee for a detailed and in-depth report that helped simplify and improve this paper. AGR acknowledges support from the Spanish grants AYA2016-79006-P and PID2019-108765GB-I00, funded by MCIN/AEI/10.13039/501100011033 and by "ERDF A way of making Europe". AU acknowledges support from the Spanish grants PGC2018-094671-B-I00, funded by MCIN/AEI/10.13039/501100011033 and by ``ERDF A way of making Europe'', and PID2019-108765GB-I00, funded by MCIN/AEI/10.13039/501100011033. FB, JP, AB, IB, and JdB acknowledge funding from the European Research Council (ERC) under the European Union’s Horizon 2020 research and innovation programme (grant agreement No.726384/Empire). 
The work of AKL is partially supported by the National Science Foundation under Grants No. 1615105, 1615109, and 1653300. TS acknowledges funding from the European Research Council (ERC) under the European Union’s Horizon 2020 research and innovation programme (grant agreement No. 694343). MC gratefully acknowledges funding from the Deutsche Forschungsgemeinschaft (DFG) through an Emmy Noether Research Group, grant number CH2137/1-1 and the DFG Sachbeihilfe, grant number KR4801/2-1. COOL Research DAO is a Decentralized Autonomous Organization supporting research in astrophysics aimed at uncovering our cosmic origins. CE acknowledges funding from the Deutsche Forschungsgemeinschaft (DFG) Sachbeihilfe, grant No. BI1546/3-1. SGB acknowledges support from the Spanish MINECO and FEDER funding grant AYA2016-76682-C3-2-P.  SCOG acknowledges support from the DFG via SFB 881 “The Milky Way System” (sub-projects B1, B2 and B8) and from the Heidelberg cluster of excellence EXC 2181-390900948 “STRUCTURES: A unifying approach to emergent phenomena in the physical world, mathematics, and complex data”, funded by the German Excellence Strategy. RSK acknowledges financial support from the German Research Foundation (DFG) via the collaborative research centre (SFB 881, Project-ID 138713538) “The Milky Way System” (subprojects A1, B1, B2, and B8). He also thanks for funding from the Heidelberg Cluster of Excellence ``STRUCTURES'' in the framework of Germany’s Excellence Strategy (grant EXC-2181/1, Project-ID 390900948) and for funding from the European Research Council via the ERC Synergy Grant ``ECOGAL'' (grant 855130). JP acknowledges support from the Programme National “Physique et Chimie du Milieu Interstellaire” (PCMI) of CNRS/INSU with INC/INP co-funded by CEA and CNES. MQ and SGB acknowledges support from the Spanish grant PID2019-106027GA-C44, funded by MCIN/AEI/10.13039/501100011033. ER acknowledges the support of the Natural Sciences and Engineering Research Council of Canada (NSERC), funding reference number RGPIN-2022-03499. ES and TGW acknowledge funding from the European Research Council (ERC) under the European Union’s Horizon 2020 research and innovation programme (grant agreement No. 694343). MCS acknowledges financial support from the German Research Foundation (DFG) via the collaborative research centre (SFB 881, Project-ID 138713538) ”The Milky Way System” (subprojects A1, B1, B2, and B8). Y-HT acknowledges funding support from NRAO Student Observing Support Grant SOSPADA-012 and from the National Science Foundation (NSF) under grant No. 2108081. 

This paper makes use of the following ALMA data: ADS/JAO.ALMA\#2018.1.01530.S. ALMA is a partnership of ESO (representing its member states), NSF (USA) and NINS (Japan), together with NRC (Canada), MOST and ASIAA (Taiwan), and KASI (Republic of Korea), in cooperation with the Republic of Chile. The Joint ALMA Observatory is operated by ESO, AUI/NRAO and NAOJ.

We also thank the staff of the IRAM ARC node at IRAM Grenoble for their support with data reduction.
\end{acknowledgements}

\bibliographystyle{aa}
\bibliography{dense_gas_excitation-v01}

\begin{appendix}

\section{Summary of line ratios}\label{app_SummlineRatios}

\begin{figure*}[!htp]
\centering
\includegraphics[width=\linewidth]{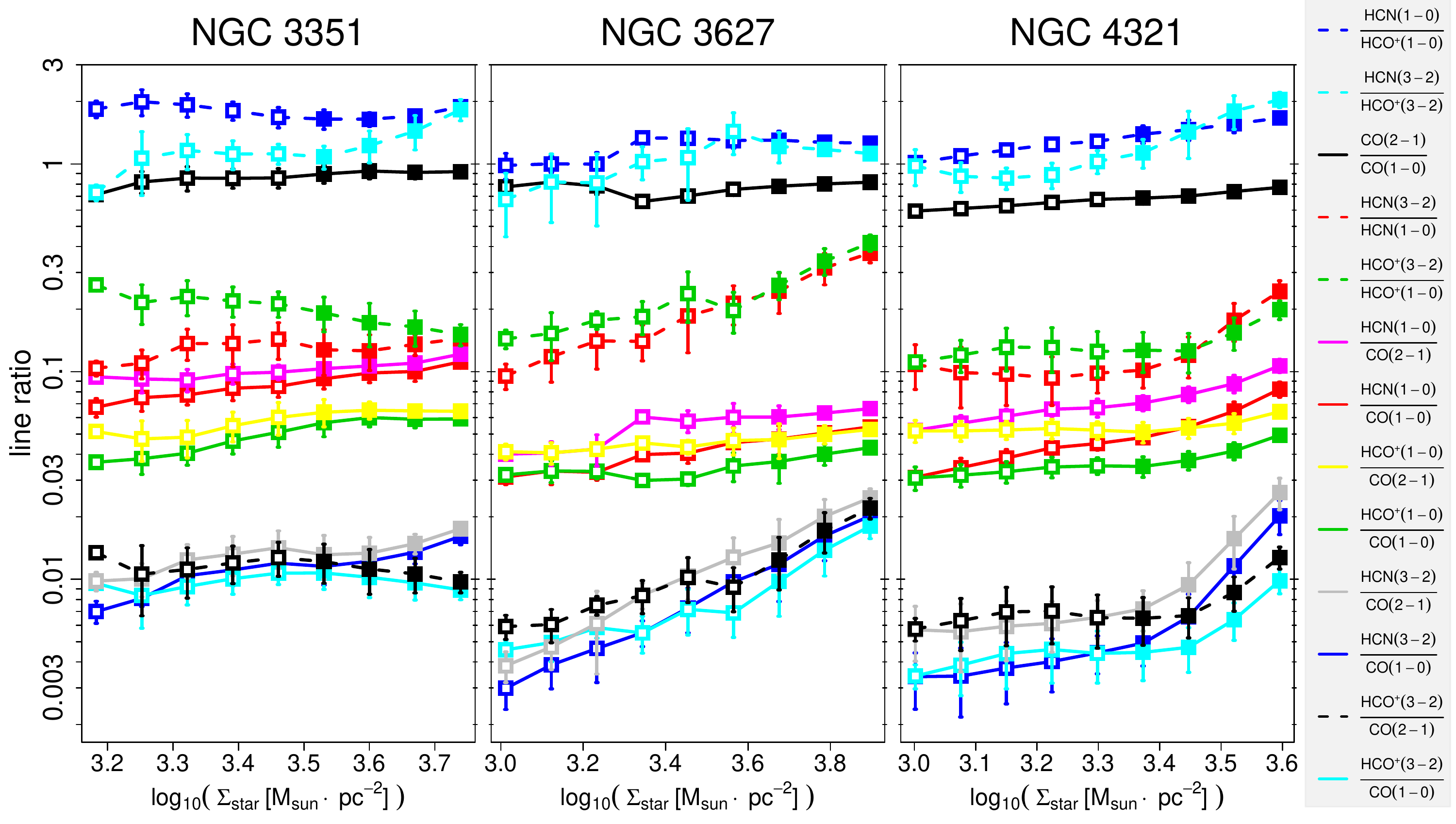}
\caption{Binned trends of the line ratios in Table~\ref{tab_lineRatios} (plus the line/\coone ones) as a function of $\sstar$. Symbols are defined as in Figure~\ref{fig_line-ratio_galaxy_comparison}.}\label{fig_line-ratio_comparison}
\end{figure*}

As a potential reference for future studies, we present here the typical values of a set of key line ratios derived from the six line maps per galaxy. Table~\ref{tab_lineRatios} summarises the percentiles of ratios measured in the S/N-selected regions in \ngc{3351}, \ngc{3627}, and \ngc{4321}. For a comprehensive view of their systematic variations across galaxy discs, we plot their binned trends as a function of $\Sigma_\textrm{star}$ in Fig.~\ref{fig_line-ratio_comparison}. Most line ratios tend to increase along with $\sstar$, with varying slopes.

\begin{table*}[!htp]
\centering
\caption{Representative line ratios within our galaxy sample}\label{tab_lineRatios}
\resizebox{\textwidth}{!}{  
\begin{tabular}{rccc@{\hspace{0.8cm}}ccc@{\hspace{0.8cm}}ccc@{\hspace{0.8cm}}ccc}
\hline
\hline
\noalign{\smallskip}
Line ratio & \multicolumn{3}{c}{\hspace{-0.8cm} Whole sample} & \multicolumn{3}{c}{\hspace{-0.8cm}\ngc{3351}} & \multicolumn{3}{c}{\hspace{-0.8cm}\ngc{3627}} & \multicolumn{3}{c}{\hspace{-0.4cm}\ngc{4321}} \\ \noalign{\smallskip}
 & 16\% & 50\% & 84\% & 16\% & 50\% & 84\% & 16\% & 50\% & 84\% & 16\% & 50\% & 84\% \\
\noalign{\smallskip}
\hline
\noalign{\smallskip}
  \cotwo/\coone & 0.682 & 0.781 & 0.912 & 0.802 & 0.912 & 0.976 & 0.723 & 0.789 & 0.838 & 0.636 & 0.701 & 0.766 \\ \noalign{\medskip} 
  \hcnone/\cotwo & 0.055 & 0.074 & 0.108 & 0.095 & 0.106 & 0.119 & 0.040 & 0.057 & 0.067 & 0.064 & 0.075 & 0.098 \\ \noalign{\medskip}
  \hcopone/\cotwo & 0.044 & 0.054 & 0.064 & 0.054 & 0.063 & 0.067 & 0.041 & 0.045 & 0.053 & 0.047 & 0.055 & 0.063 \\ \noalign{\medskip} 
  \hcnthree/\cotwo & 0.006 & 0.012 & 0.019 & 0.011 & 0.015 & 0.017 & 0.005 & 0.012 & 0.023 & 0.006 & 0.009 & 0.022 \\ \noalign{\medskip}
  \hcopthree/\cotwo & 0.006 & 0.009 & 0.014 & 0.009 & 0.011 & 0.014 & 0.007 & 0.010 & 0.020 & 0.005 & 0.007 & 0.011 \\ \noalign{\medskip} 
  \hcnthree/\hcnone & 0.099 & 0.143 & 0.250 & 0.105 & 0.141 & 0.156 & 0.115 & 0.205 & 0.340 & 0.089 & 0.121 & 0.214 \\ \noalign{\medskip}
  \hcopthree/\hcopone & 0.122 & 0.175 & 0.258 & 0.140 & 0.180 & 0.233 & 0.156 & 0.222 & 0.371 & 0.105 & 0.137 & 0.186 \\ \noalign{\medskip} 
  \hcnone/\hcopone & 1.186 & 1.401 & 1.761 & 1.587 & 1.746 & 1.936 & 0.971 & 1.234 & 1.337 & 1.234 & 1.406 & 1.681 \\ \noalign{\medskip} 
  \hcnthree/\hcopthree & 0.872 & 1.164 & 1.746 & 0.993 & 1.218 & 1.789 & 0.677 & 1.109 & 1.293 & 0.909 & 1.249 & 2.040 \\ \noalign{\smallskip}
\hline
\end{tabular}}
\tablefoot{For each line ratio indicated in the leftmost column, we show the 16$^\mathrm{th}$, 50$^\mathrm{th}$, and 84$^\mathrm{th}$ percentiles of values measured on a pixel-by-pixel basis in the whole sample of galaxies and in each galaxy separately. In all cases, only pixels where S/N$\geq5$ for all the lines are considered.}
\end{table*}

\begin{figure*}[!ht]
\includegraphics[trim= 0 0 0 0, clip,width=\linewidth]{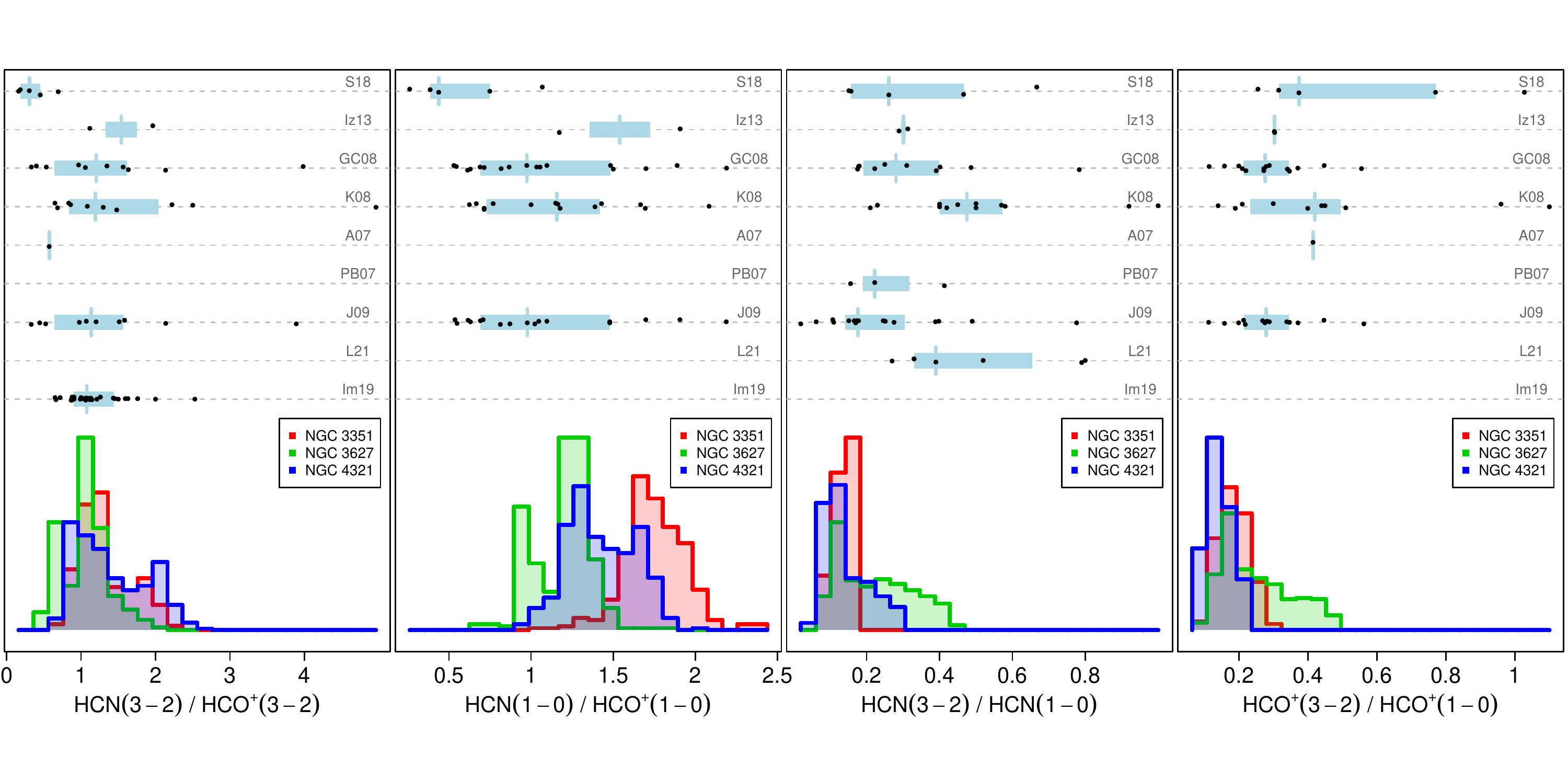} 
\caption{Histograms per galaxy of four line ratios as in Fig.~\ref{fig_litHist-simple}. Here we show literature data as black dots for comparison. In each row, the label corresponds to the paper from which the measurements are taken (see text for details), while a light-blue shaded bar indicates the 25$^\mathrm{th}$-75$^\mathrm{th}$ percentile range of the data set and a vertical segment indicates the median.}\label{fig_litHist}
\end{figure*}

To put our spatially resolved measurements of  \hcnthree and \hcopthree into context, we compare them in Fig.~\ref{fig_litHist} with observations from the literature, dominated by AGN and starburst galaxies. The histograms representing our three target galaxies are the same as in Fig.~\ref{fig_litHist-simple}. Here we add black dots that correspond to measurements from a heterogeneous (in terms of samples and spatial resolution) but comprehensive set of studies. 
Single-dish, spatially unresolved ratios are taken from \citet[labelled GC08]{gra08}, \citet[K08]{kri08}, \citet[A07]{aal07}, \citet[PB07]{per07}, and \citet[J09]{jun09}. Spatially resolved ratios are taken from \citet[Iz13; AGN and starburst peaks of \ngc{1097} from their Table~8]{izu13}, \citet[S18; five 3\si{\arcsec}-apertures in the LIRG VV\,114 from their Table~2]{sai18}, \citet[Im19; 25 (U)LIRG centres at $\sim$500-pc resolution]{ima19}, and \citet[L21; seven regions of three (U)LIRGs from their Table 4]{led21}. There is a clear contrast between the fixed-transition HCN-to-HCO$^+$ ratios (two leftmost panels) and the fixed-species 3--2/1--0 ratios (two rightmost panels). In the former two, the overlap between our data and the literature values is significant. In the latter two, our 3--2/1--0 data clearly cluster at the lower range of the literature. 

\end{appendix}
\end{document}